\documentclass[a4paper,fleqn,usenatbib]{mnras}

\usepackage[T1]{fontenc}
\usepackage{ae,aecompl}
\usepackage{pdflscape}

\usepackage{graphicx}	% Including figure files
\usepackage{amsmath}	% Advanced maths commands
\usepackage{amssymb}	% Extra maths symbols
\usepackage{long table}	% Long table

\newcommand{\pc}{\,pc\,}
\newcommand{\pcw}{\,pc}
\newcommand{\kpc}{\,kpc\,}
\newcommand{\kpcw}{\,kpc}

\title[Testing {\sevensize PARSEC} v1.2S]{Testing Models of Stellar Structure and Evolution I.
Comparison with Detached Eclipsing Binaries
}

\author[C. del Burgo and C. Allende Prieto]{
C. del Burgo$^{1}$\thanks{E-mail: cburgo@inaoep.mx}
and C. Allende Prieto$^{2,3}$
\\
$^{1}$Instituto Nacional de Astrof\'{\i}sica, \'Optica y Electr\'onica, Luis Enrique Erro 1, Sta. Ma. Tonantzintla, Puebla, Mexico\\
$^{2}$Instituto de Astrof\'{\i}sica de Canarias, 38205 La Laguna, Tenerife, Spain\\
$^{3}$Departamento de Astrof\'{\i}sica, Universidad de La Laguna, 38206 La Laguna, Tenerife, Spain
}

\date{Accepted XXX. Received YYY; in original form ZZZ}

\pubyear{2018}

\begin{document}
\label{firstpage}
\pagerange{\pageref{firstpage}--\pageref{lastpage}}
\maketitle

\begin{abstract}
%top-notch
We present the results of an analysis aimed at testing the accuracy and precision of the PARSEC v1.2S library of stellar evolution models, in a Bayesian framework, 
to infer stellar parameters. We mainly employ the online DEBCat catalogue by Southworth, a compilation of detached eclipsing binary systems with published 
measurements of masses and radii to $\sim$ 2 per cent precision. We select a sample of 318 binary components, with masses between 0.10 and 14.5 M$_{\sun}$, 
at distances between 1.3 \pc and $\sim$ 8 \kpc for Galactic objects and $\sim$ 44--68 \kpc for extragalactic ones. The Bayesian analysis applied takes as input 
effective temperature, radius, and [Fe/H], and their uncertainties, returning theoretical predictions for other stellar parameters. From the comparison with 
dynamical masses, we conclude that the inferred masses are precisely derived for stars on the main-sequence and in the core-helium-burning phase, with 
uncertainties of 4 per cent and 7 per cent, respectively, on average. Masses for subgiants and red giants are predicted within 14 per cent, and those for early 
asymptotic giant branch stars within 24 per cent. These results are helpful to further improve the models, in particular for advanced evolutionary stages for which 
our understanding is limited. We obtain distances and ages for the binary systems and compare them, whenever possible, with precise literature estimates, 
finding excellent agreement. We discuss evolutionary effects and challenges for inferring stellar ages from evolutionary models. We also provide useful 
polynomial fittings to theoretical zero-age main-sequence relationships.
\end{abstract}

\begin{keywords}
stars: fundamental parameters -- stars: evolution -- stars: statistics -- (stars:) binaries: eclipsing -- (stars:) Hertzsprung-Russell and colour-magnitude diagrams
\end{keywords}

%%%%%%%%%%%%%%%%%%%%%%%%%%%%%%%%%%%%%%%%%%%%%%%%%%

%%%%%%%%%%%%%%%%% BODY OF PAPER %%%%%%%%%%%%%%%%%%

\section{Introduction}

The analysis of integrated light from stellar populations or entire galaxies heavily relies on having accurate stellar structure and evolution models.
Such theoretical calculations for single stars bring forth predictions of their fundamental parameters as a function of evolutionary stage.
Models are increasingly sophisticated, but still suffer from well-known shortcomings related to incomplete or oversimplified
physics: mixing processes, radiative opacities, or nuclear reaction rates, to mention a few examples.
The comparison between theory and observations has revealed inadequacies, which are more acute for giants, low mass red dwarfs, and  
pre-main-sequence stars. We refer the reader to the extensive literature about 
stellar evolution models \citep[][etc.]{bressan2012,chabrier2000,dotter2008,meynet2000,pietrinferni2004,yi2001}.

Uncertainties in stellar parameters derived from theoretical models have a serious impact on our understanding of the formation and evolution of the Galaxy.
They propagate into systematic uncertainties in the initial mass function, the age-activity-rotation relation, and the inferred exoplanet masses and radii.
Alternatively, one can use empirical relations among stellar parameters calibrated on the Sun, binaries, or stellar populations with reliable parameters.
It is very important to significantly expand the databases of accurate parameters of stars belonging to stellar clusters and binary systems, 
in order to test theoretical predictions.

A detached eclipsing binary (DEB) system consists of two non-interacting stars that have evolved as if they were single 
and whose orbital plane is nearly or perfectly aligned towards the observer, so this can observe periodic eclipses. 
These systems are very important in Astronomy, since it is possible to directly determine the masses and radii of their components with very high accuracy. 
Some of them are composed of stars that have left the main-sequence, e.g. subgiants and red giants. 
These are particularly useful to test stellar evolution calculations of evolutionary stages for which our understanding is incomplete. 
Systems where the components are in different evolutionary stages are also very important  to test theoretical models.

ESA's \textit{Hipparcos} mission \citep[][]{perryman2009} supplied accurate stellar parallaxes that reduced the uncertainties on the Hertzsprung--Russell diagram (HRD)
and the colour-magnitude diagram (CMD) for nearby field stars, and consequently it yielded more precise comparisons with stellar evolution calculations. 
Despite degeneracies between mass, age, and metallicity, Hipparcos data made it possible to estimate masses, radii, and effective temperatures 
with uncertainties of $\simeq$ 8, 6 and 2 per cent, respectively, at least for solar metallicity stars \citep[][]{allendeprieto1999}.
ESA's \textit{Gaia} mission \citep[][]{gaia2016, gaia2018} is currently collecting astrometric, photometric, and spectroscopic data for $\sim$ 10$^9$ stars 
with much higher precision and sensitivity than its predecessor, to further constrain the stellar parameters and immensely increase the stellar census of the Galaxy. 

In this paper we present a study of the accuracy and precision of a state-of-the-art library of stellar evolution models. 
We apply a Bayesian method to infer stellar parameters and check the consistency between the derived masses and accurate dynamical values from 
the literature for carefully selected DEB stars. 
We further investigate the precision of our procedure to yield stellar masses by means of simulations.
We map the differences between input and output values of stellar mass and age for a collection of models on the CMD, and analyse the impact 
of degeneracies on the parameters determination. We provide distances and discuss issues related to the prediction of stellar ages. 
Section \ref{evolutionmodels} outlines the stellar evolution models and defines the original grid used in this study.  
Section \ref{binaries} describes the selected samples of DEB stars.
Section \ref{method} explains the methodology applied to derive stellar parameters.
The analysis and results are given in Section \ref{results}. 
Section \ref{discussion} contains a summary and a discussion of our findings. 
In Section \ref{conclusions} we outline our conclusions.
Appendix \ref{appendix:grid} presents the interpolated grid of models that we also utilised for this work.
In Appendix \ref{appendix:evolution} a series of stellar evolution effects are illustrated.
Appendix \ref{appendix:likelihood} shows examples of likelihood functions for two DEB systems.
Finally, Appendix \ref{appendix:zams} provides polynomial fittings to theoretical zero-age main-sequence parameters at different metallicities.

\section{Stellar evolution models}
\label{evolutionmodels}

The {\sevensize PARSEC} (the PAdova and TRieste Stellar Evolution Code) models provide isochrones and stellar tracks  
that follow the evolution of a star from its formation to the asymptotic giant branch phase for low- and intermediate-mass stars, 
or the carbon ignition phase for massive stars. 
The library distinguishes between the following stellar evolutionary phases: pre-main-sequence (PMS), 
main-sequence (MS), sub-giant branch (SGB), red giant branch (RGB), 
different stages of the core helium burning phase (CHeB), early asymptotic giant branch (EAGB), 
and thermally pulsing asymptotic giant branch (TP-AGB).
The library does not include stellar remnants such as white dwarfs.

Stars spend most of their lives burning hydrogen into helium on the MS phase.
The lifetime of a star ranges from a few million years for very massive stars to ages significantly longer than that of the Universe for very low mass red dwarfs. 
However, the maximum theoretical age of the models used here is 13.5 Gyr. 

The {\sevensize PARSEC} models provide, among other stellar parameters, the actual mass\footnote{Also denoted by $M_\text{m}$ when 
it is compared with the dynamical mass of an eclipsing binary star.} ($M$), luminosity ($L$), effective temperature ($T_\text{eff}$), 
surface gravity ($g$), bolometric magnitude ($M_\text{bol}$), and magnitudes in a chosen photometric system  
as a function of age ($\tau$), initial metallicity ($Z_\text{ini}$), and initial mass ($M_\text{0}$).
The radius ($R$) can be trivially calculated from the mass and gravity, and the mean density ($\rho$) from the mass and radius.

Similarly to \citet{delburgo2016}, we adopted the nominal values of the International Astronomical Union (IAU) 2015 
Resolution B3 \citep[][]{prsa2016}: $T_{\text{eff},{\sun}}$=5772\,K for the solar effective temperature, 
GM$_{\sun}$=1.3271244 10$^{20}$ m$^3$ s$^{-2}$ for the product of the gravitational constant and the solar mass, 
L$_{\sun}$=3.828 10$^{33}$ erg s$^{-1}$ for the solar mean radiative luminosity, and R$_{\sun}$=6.957 10$^8$ m for the solar radius.
We employed G=6.67428 10$^{-11}$ m$^3$ kg$^{-1}$ s$^{-2}$, which is recommended 
by the IAU Working Group on Numerical Standards for Fundamental Astronomy, NSFA \citep[][]{luzum2011}.
We updated the masses and luminosities of the models, and then the effective temperatures from the radii and luminosities, 
applying the relationship $T_\text{eff}/T_{\text{eff},{\sun}}=\left(R/\text{R}_{\sun}\right)^{1/2} \left(L/\text{L}_{{\sun}}\right)^{1/4}$. 
The differences with respect to the tabulated values in the {\sevensize PARSEC} library are negligible anyway. 
We also computed the initial mass function from the relation of \citet{chabrier2001} (see Section \ref{method}).

\begin{figure*}
\centering
 \includegraphics[width=175mm,angle=0]{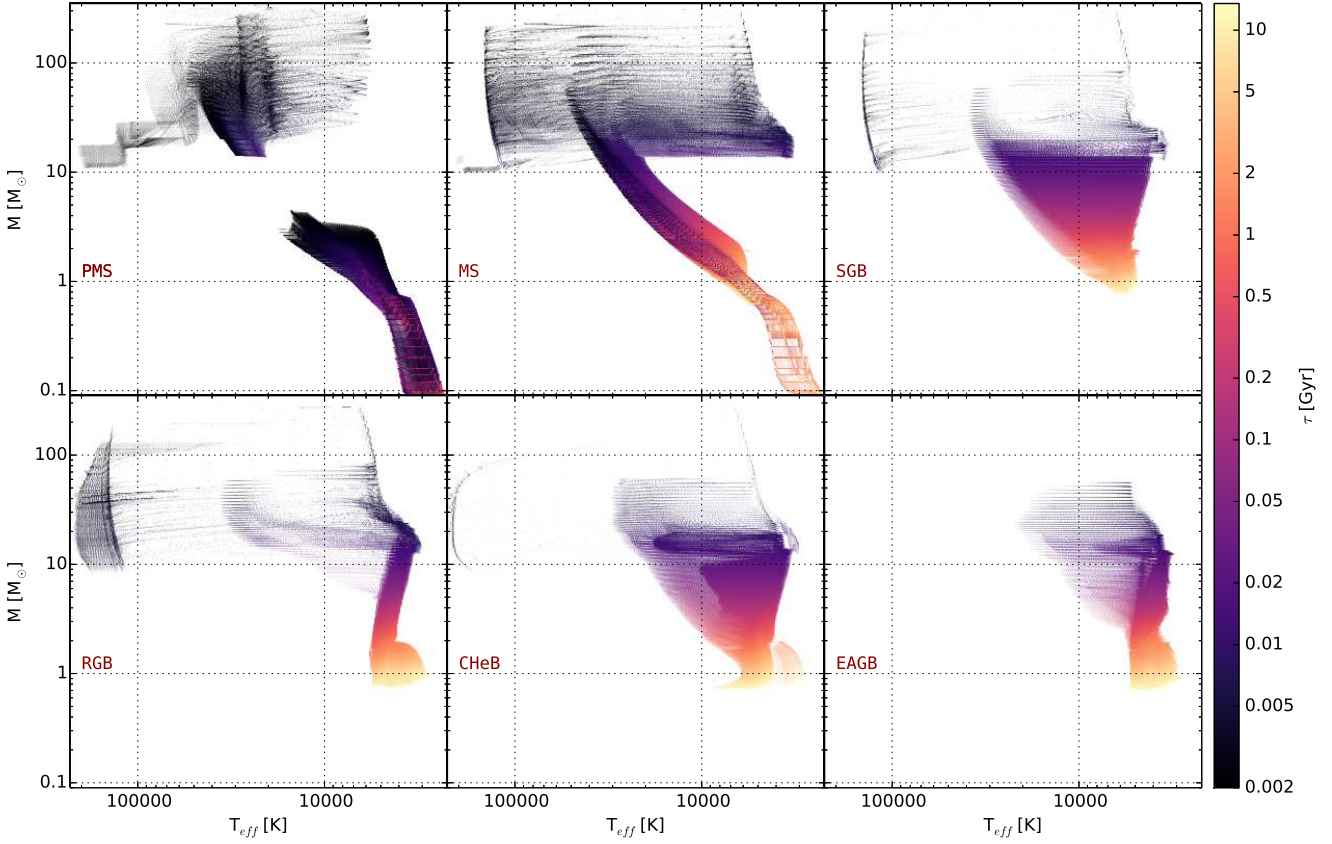}
 \caption{Loci of the {\sevensize PARSEC} v1.2S models, colour-coded by age, 
               in the $T_\text{eff}$--$M$ diagram for the stellar evolutionary stages PMS, MS, SGB, RGB, CHeB, and EAGB.
               Each point represents a model from the original values in the library, before regridding. 
               Note the age ruler by {\sevensize PARSEC} calculations includes the PMS lifetime of the stars.
}
 \label{fig:fig1}
\end{figure*}

We downloaded\footnote{http://stev.oapd.inaf.it/cgi-bin/cmd} and arranged a grid 
of {\sevensize PARSEC} Isochrones \citep[version 1.2S,][]{bressan2012, chen2014, chen2015, tang2014}, 
where the iron-to-hydrogen ratio, [Fe/H]\footnote{These models assume solar-scaled metal abundances and for their choice of 
the solar mixture the metal mass fraction can be computed as $Z$=0.01524 10$^{\text{[Fe/H]}}$.}
ranges from -2.15 to -0.95 in steps of 0.05 dex, from -0.95 to -0.65 in steps of 0.03 dex, and from -0.65 to 0.42 in steps of 0.01 dex. 
The age $\tau$ spans from 2 Myr to 13.5 Gyr in steps of 5 per cent.  
The initial mass $M$ runs from 0.09 M$_{\sun}$ to the highest mass established by the stellar lifetimes. 
The absolute maxima for the initial mass and actual mass in the grid are 350.0 M$_{\sun}$ and 345.2 M$_{\sun}$, respectively.

The {\sevensize PARSEC}'s original grid fairly well samples the initial mass for all evolutionary phases, although in irregular steps.
We resampled the 6 133 173 models, grouped in 25 521 isochrones, to a resolution of 0.1 per cent in mass using linear interpolation.
The resulting interpolated grid has a greater number (105 853 725) of models than the original one, but the number density distribution in 
the $T_\text{eff}$--$M$ diagram significantly changes across the evolutionary stages, with lower representations towards 
later phases (see Appendix \ref{appendix:grid}). 
We employed the original and interpolated grids to infer the stellar parameters. 
The irregularly spaced grid was adopted when the interpolated grid provides a lower number of models.
The {\sevensize PARSEC} v1.2S isochrones are carefully interpolated from stellar evolutionary tracks \citep[][]{bressan2012}. 
This is a complex procedure subject to systematic errors, and which needs close monitoring of the associated uncertainties \citep[see e.g.,][]{dotter2016}. 
In this work we adopt the isochrones provided with the {\sevensize PARSEC} v1.2S models and trust that they are appropriate for our purposes.

\begin{figure}
\centering
 \includegraphics[width=85mm,angle=0]{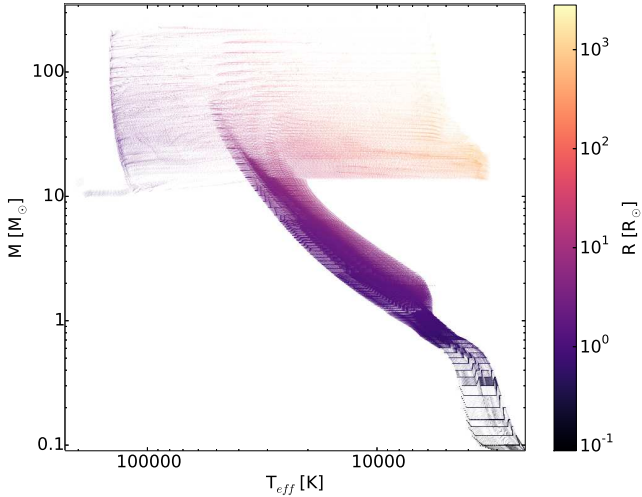}
 \caption{Loci of the {\sevensize PARSEC} v1.2S models, colour-coded by radius, 
               in the $T_\text{eff}$--$M$ diagram for the main-sequence.
               Each point represents a model from the original values in the library, before regridding.
}
 \label{fig:fig2}
\end{figure}

\begin{figure}
\centering
 \includegraphics[width=85mm,angle=0]{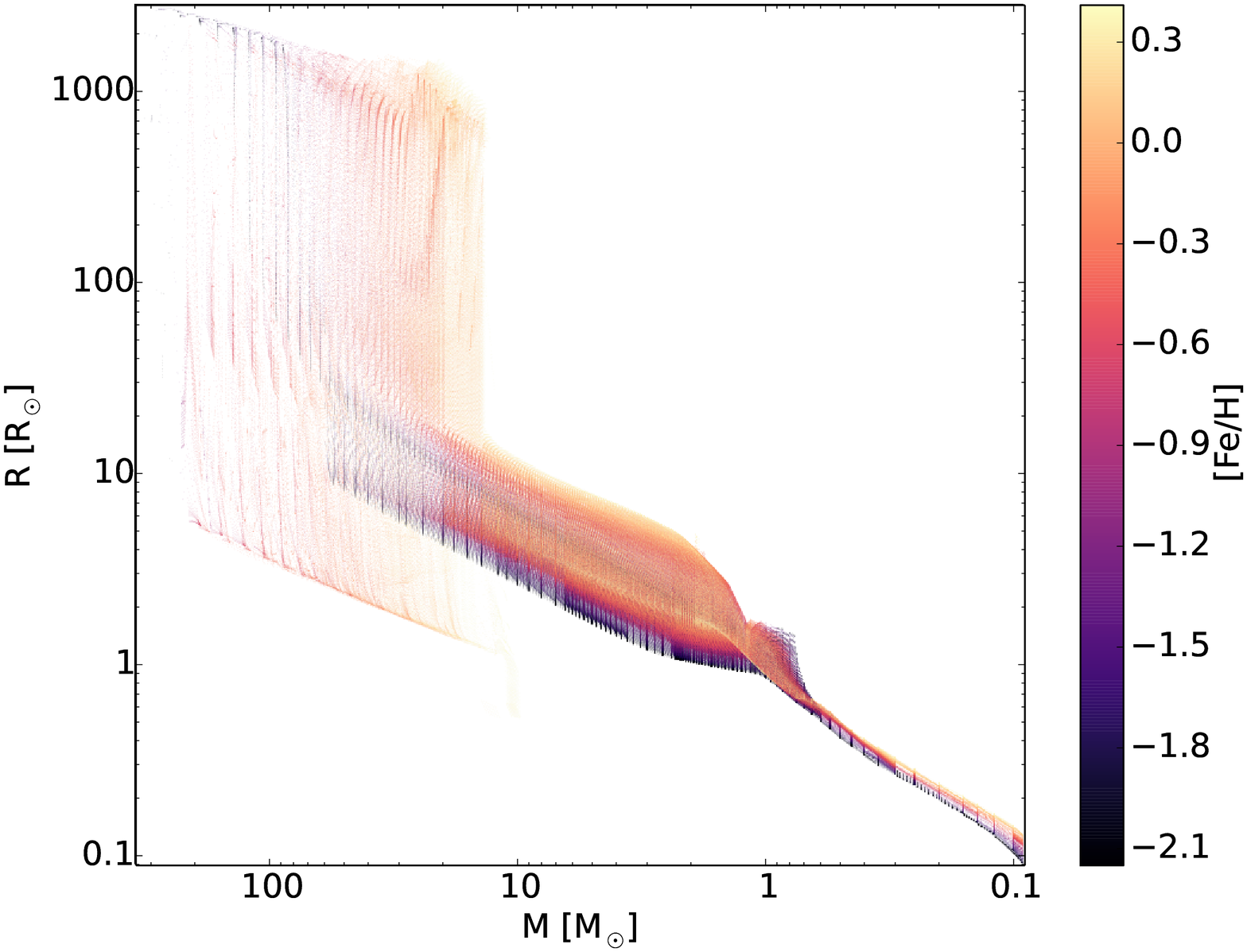}
 \caption{Loci of the {\sevensize PARSEC} v1.2S models, colour-coded by [Fe/H], 
               in the $M$--$R$ diagram for the main-sequence.
               Each point represents a model from the original values in the library, before regridding.
}
 \label{fig:fig3}
\end{figure}

This research was aimed at testing the {\sevensize PARSEC} v1.2S models to infer stellar parameters.
We inspected all the evolutionary stages distinguished by the library apart from the TP-AGB stage.
Note the study of the pre-main-sequence evolution includes more complex physics and its modelling is arduous, particularly 
for objects with masses under 1 M$_{\sun}$ \citep[see e.g., ][]{stassun2014}.

Figure \ref{fig:fig1} shows the theoretical stellar age against $T_\text{eff}$ and $M$, from the {\sevensize PARSEC} v1.2S library. 
Stars at different evolutionary stages may have similar effective temperatures and masses.
In addition, there are models for the same evolutionary stage with a wide range of ages in similar positions of the diagram $T_\text{eff}$--$M$. 
Figures \ref{fig:fig2} and \ref{fig:fig3} respectively display the theoretical radius against $T_\text{eff}$ and $M$ and [Fe/H] against $M$ and $R$ for the MS stage. 

We stress that the {\sevensize PARSEC} v1.2S models use a solar composition with Z = 0.01524 and assume all metals are scaled in the same proportions 
for stars with higher or lower metal abundances \citep[see][]{bressan2012}. 
This is appropriate for stars with near-solar metallicities, as usually found in the Galactic disc, but it is not for metal-poor stars, 
which usually exhibit enhancements in the $\alpha$ elements (O, Ne, Mg, Si, S, Ca, Ti). 
Nevertheless, we find excellent agreement between the inferred masses and dynamical masses for metal-poor stars too, 
and the same is true for the ages of the systems in globular clusters (see Section \ref{results}), which indicates that the impact of the 
$\alpha$-element enhancement in the parameters derived following our methodology is overall quite limited.

\section{Detached eclipsing binary stars}
\label{binaries}

There is a number of compilations of reliable measurements for the stellar parameters of DEB stars in the literature. 
We compared two of them, namely \citet{torres2010} and the online DEBCat catalogue\footnote{http://www.astro.keele.ac.uk/jkt/debcat/} by \citet{southworth2015}, 
in order to select the most appropriate inputs for this research.

\begin{figure}
\centering
 \includegraphics[width=85mm,angle=0]{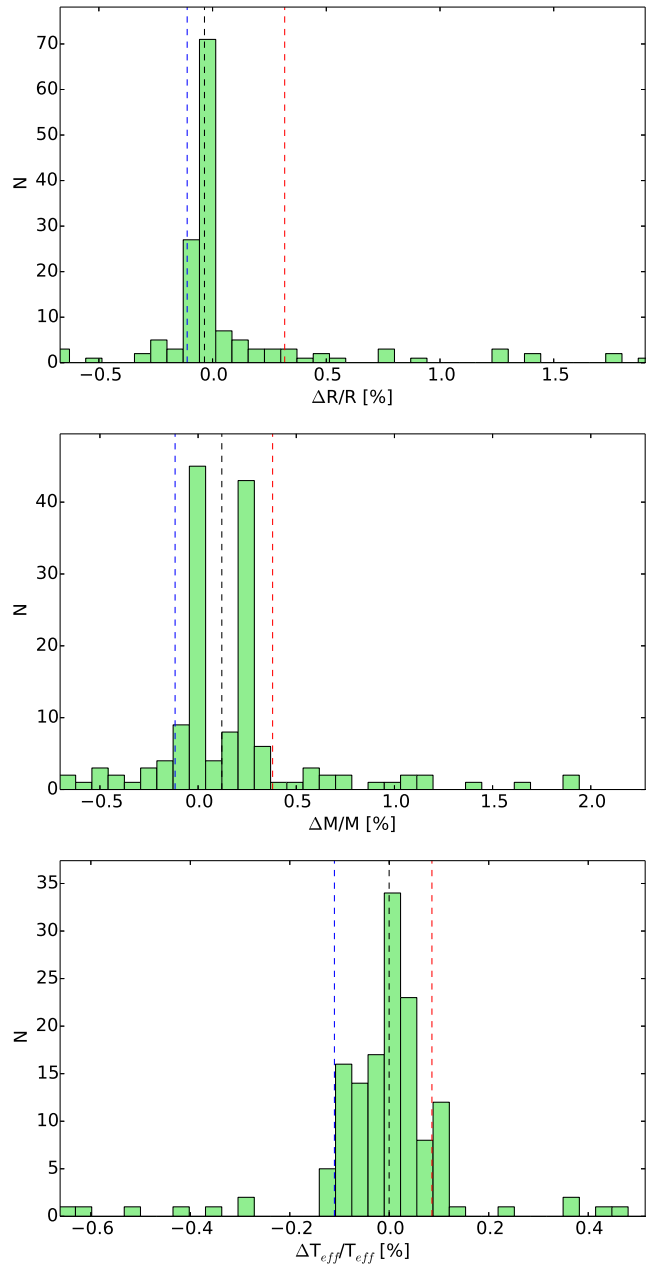}
 \caption{Relative differences in radius, mass and effective temperature for the binary stars in the 84 systems in common in the compilations 
 from \citet{southworth2015} and \citet{torres2010}.
 Vertical dashed lines in black, blue, and red correspond to the median and the 34th percentiles on the left and the right sides, respectively.  
}
 \label{fig:fig4}
\end{figure}

\subsection{Description of the DEB catalogues}

\citet{torres2010} compiled accurate measurements of masses and radii for a sample of 94 DEB systems and $\alpha$ Centauri (astrometric binary) to 3 per cent precision.
The sum of masses scales with the third power of the sum of the radial velocity amplitudes. 
Thus, an accurate constraint on the orbital velocities leads to accurate masses. That is possible from high-resolution, high-signal-to-noise spectra. 
Spectroscopic information combined with light curve analysis provide robust determinations of orbital parameters and stellar radii for the components of DEB systems. 
\citet{torres2010} consistently recomputed literature parameters for the sample of 95 binary systems and added others such as the amount of 
interstellar extinction, $T_\text{eff}$, and [Fe/H] when available.
It mostly consists of DEB systems in our Galaxy at distances from 1.3\pc to 2.8\kpc and includes an extragalactic binary at 51\kpc. 
The number of binary systems with spectroscopic metallicities in their sample is 21.

DEBCat \citep[][]{southworth2015} is a catalogue of 199 DEB systems with measurements for the mass and radius to 2 per cent precision for most of them. 
It is regularly updated with new binaries or revised measurements. 
It presents 84 binary systems in common with \citet{torres2010}.
DEBCat collects, among other stellar parameters, the mass, radius, effective temperature, luminosity, and metallicity when available. 
It does not provide distances for the DEB systems, but it contains a number of extragalactic objects.
There are 82 systems with known metallicities, which were constrained from abundance analyses of high-resolution spectra or, 
for those belonging to a stellar cluster, may be from other cluster members. 

Despite the parameters for some binary systems have been updated with more precise measurements, \citet{torres2010} performed a methodical compilation. 
When possible, they updated the interstellar reddening estimates to derive effective temperatures.
Their adopted value for GM$_{\sun}$ matches the nominal value of the International Astronomical Union (IAU) 2015 Resolution B3 \citep[][]{prsa2016}.
We updated their radii to be consistent with the IAU 2015 Resolution B3. Nevertheless, they used R$_{\sun}$=6.9566 10$^8$ m, 
a solar radius that is negligibly smaller (by 0.006 per cent) than the nominal value adopted in the resolution.

\subsection{Comparison of the DEB catalogues}
\label{comparingDEBcat}

Figure \ref{fig:fig4} displays the relative differences (expressed as percentages) in radius ($\Delta R/R$), mass ($\Delta M/M$), 
and effective temperature ($\Delta T_\text{eff}/T_\text{eff}$) from DEBCat with respect to the catalog of \citet{torres2010}.
Note $\Delta R/R$ and $\Delta T_\text{eff}/T_\text{eff}$ are single-peak distributions, while $\Delta M/M$ has two peaks, one around 0 and the other around 0.25 per cent.
$\Delta T_\text{eff}/T_\text{eff}$ is very narrow, with a median and 34th percentiles on the left and right sides of -0.00029, -0.11, and 0.09 per cent, respectively. 
The median and 34th percentiles on the left and right sides of $\Delta R/R$ are -0.036, -0.11, and 0.32 per cent, respectively.
Finally, the median and 34th percentiles on the left and right sides of $\Delta M/M$ are respectively 0.12, -0.12, and 0.38 per cent. 
The median of $\Delta M/M$ is between the two peaks, which are enclosed by the 34th percentiles. 
We conclude the two collections are in very good agreement. The observed discrepancies are due to the use of different datasets, solar units, and methods. 
For example, there are different $T_\text{eff}$ scales.
We note, however, the impact of employing distinct values for M$_\odot$ and R$_\odot$ in DEBCat is negligible. 

\begin{figure}
\centering
 \includegraphics[width=85mm,angle=0]{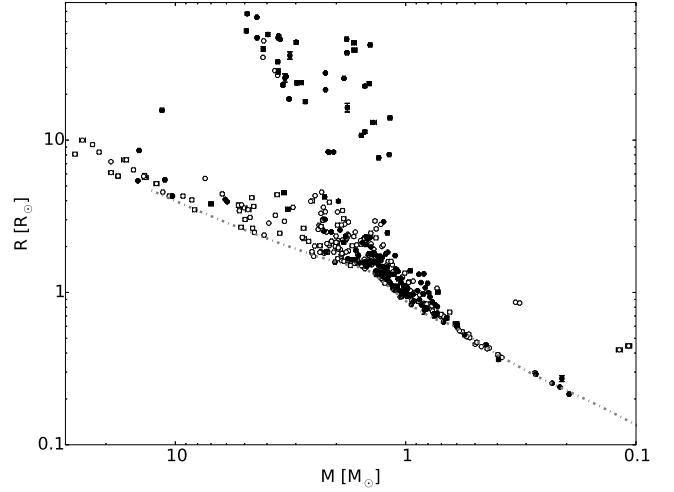}
 \caption{Radius versus mass for detached eclipsing binaries. 
 There are 199 DEB systems from DEBCat \citep[][]{southworth2015} and two more systems with metallicities 
 from \citet{torres2010} that we added to our selection ($\alpha$ Cen is not shown). 
 Open and filled symbols stand for objects with and without metallicities (see Sect. \ref{selectionfeh}), respectively. 
 Uncertainties are mostly smaller than the plotted symbols.
 A theoretical ZAMS for solar metallicity from {\sevensize PARSEC} v1.2S models is marked by the gray dash-dotted line.
}
 \label{fig:fig5}
\end{figure}

DEBCat contains nearly 4 times more objects with metallicities than \citet{torres2010}.
It singularly includes a significant number of giant stars and some low mass red dwarfs with metallicities, conversely to the catalogue of \citet{torres2010}, with only a few. 
Figure \ref{fig:fig5} plots $R$ vs. $M$, distinguishing between stars with and without metallicities.
It is worth noting that, as result of stellar evolution, stars move up in the diagram after the end of the main sequence. 
The theoretical zero-age main-sequence (ZAMS) for solar metallicity from {\sevensize PARSEC} v1.2S models is shown as a reference.

Given the commendable agreement between the two compilations but the significant greater number of systems in DEBCat, we selected it and added 
some additional information to it. We employed the resulting database as a proxy for the DEB stars to test {\sevensize PARSEC} v1.2S models.

\subsection{Sample I: selected detached eclipsing binaries with known metallicities}
\label{selectionfeh}

Our results are mainly based on the comparison between theoretical and dynamical masses for a sample of DEB stars with known metallicities, hereafter Sample I.
This mostly consists of DEBCat binaries, although we added a few others from the literature.
In particular, \citet{torres2010} includes some binary systems which are useful for our research.
It also lists the spectral types and luminosity classes of the systems V432 Aur, V568 Lyr, V578 Mon, V785 Cep, 
and the extragalactic source OGLE-LMC-ECL09114, all of them in DEBCat. 
In addition, there are some missing metallicities in DEBCat that we found in the literature. 

We noted that DEBCat contains a value for the metallicity of RW Lac, but no uncertainty. 
We assumed it to be 0.19 dex, which is the standard deviation in metallicity for DEB stars of Sample I 
that are closer than 300\pcw\footnote{We adopted this distance as the corresponding scale height for the thin disc.}. 
There are other four systems, which are V578 Mon, V453 Cyg, CV Vel, and V906 Sco, 
with no reported metallicities in DEBCat, but with tabulated values in \citet{torres2010}. 
These authors also include three different systems, which are V570 Per, V505 Per, and $\alpha$ Cen, 
with metallicities. All these seven systems were included in our selection. 
The inclusion of the binary systems V578 Mon, V453 Cyg, and CV Vel is important since 
they represent the most massive main-sequence stars in the final sample, together with V380 Cyg. 
We updated the parallax as well as the masses and radii of $\alpha$ Cen \citep[][]{pourbaix2016}.
We adopted 0.2 dex for the uncertainty in [Fe/H] for OGLE SMC113.3 4007.

In addition, we found metallicities obtained from high resolution spectroscopy for RR Lyn, which has no metallicity in DEBCat.
\citet{burkhart1991} determined that RR Lyn has [Fe/H]=0.35, and we adopted an uncertainty of 0.19 dex because this system is closer than 300\pcw.

We also aimed at increasing the small number of M-type dwarfs in this research.
Thus, we rescued YY Gem, which is part of the Castor sextuple system, including the visual pair Castor AB, 
with [Fe/H] = 0.1$\pm$0.2 \citep[see][and references therein]{torres2002}.
We used this conservative [Fe/H] uncertainty of 0.2 dex for YY Gem despite there are recent publications with more precise values of [Fe/H] for Castor AB. 
\citet{feiden2014} conclude that CM Dra may have an [Fe/H] about 0.2 dex higher than that given 
by \citet{terrien2012} (i.e., the value listed in DEBCat), with [$\alpha$/Fe] = 0.2 dex. 
Thus, we adopted [Fe/H] = -0.10 $\pm$ 0.12 dex for CM Dra.

After including the additional information, we have a total of 90 DEB systems, plus $\alpha$ Centauri, with known metallicities and accurate parameters for Sample I.
The typical relative uncertainties for $R$, $M$, and $T_\text{eff}$ are 0.9, 0.7, and 1.8 per cent, respectively. 
The typical uncertainty in [Fe/H] is 0.10 dex.

Regarding the radius, only OGLE-LMC-ECL10567 has components with relative uncertainties greater than 3 per cent, being 6 per cent for both stars.
The secondaries of PTFEB 132.707+19.810, WOCS 12009, OGLE-LMC-ECL01866, 
OGLE-LMC-ECL03160\footnote{Note we use the same abbreviations than those in DEBCat for the detached eclipsing binaries in 
the Large and Small Magellanic Clouds.}, and KIC 8430105 have relative uncertainties of 4, 4, 4, 6, and 6 per cent, respectively.
Concerning mass, there is no object with relative uncertainties greater than 3 per cent.
There are only three systems with relative uncertainties in the effective temperatures of both components larger than 3 per cent, 
namely KIC 9246715 (4 and 4 per cent), V906 Sco (5 and 5 per cent) and V1229 Tau (6 and 5 per cent).
The primary of BG Ind has a relative uncertainty of 4 per cent.
The secondaries of Tyc 5227-1023-1 and V380 Cyg present relative uncertainties of 4 and 5 per cent, respectively.

In summary, 78 per cent of 182 stars in Sample I have relative uncertainties in radius $\leq$ 1 per cent, 91 per cent are $\leq$ 2 per cent, and 96 per cent $\leq$ 3 per cent.
Then, 86 per cent of these stars present masses accurate to 1 per cent precision, 99 per cent to 2 per cent, and all of them to 3 per cent.
For effective temperature, 24 per cent are determined to 1 per cent precision, 76 per cent to 2 per cent, and 95 per cent to 3 per cent.

\subsection{Sample II: selected nearby detached eclipsing binaries with assumed solar metallicity}
\label{selectionnofeh}

We selected a second sample of DEB stars to complement our study.
For those detached eclipsing binaries with unknown metallicity that are at distances $d \lesssim$ 300\pc we adopt [Fe/H] = 0 and $\sigma$([Fe/H]) = 0.19 dex; 
the latter is the standard deviation obtained for the binaries with known metallicities that are closer than 300\pcw. 
These stars likely belong to the thin disc of the Galaxy, for which our assumption about solar metallicity is reasonable.

The resulting sample (hereafter Sample II) consists of 136 binary stars, 
with typical relative uncertainties for $R$, $M$, and $T_\text{eff}$ of 1.0, 0.8, and 2.3 per cent, respectively. 
With regard to radius, 71 per cent of stars in Sample II have relative uncertainties $\leq$ 1 per cent, 94 per cent are $\leq$ 2 per cent, and 99 per cent $\leq$ 3 per cent.
Concerning mass, 79 per cent of stars have measurements to 1 per cent, 99 per cent to 1 per cent, and all of them to 3 per cent.
Finally, 12 per cent of Sample II have relative uncertainties $\leq$ 1 per cent, 51 per cent $\leq$ 2 per cent, and 79 per cent $\leq$ 3 per cent in the case of effective temperature.

\begin{figure}
\centering
 \includegraphics[width=85mm,angle=0]{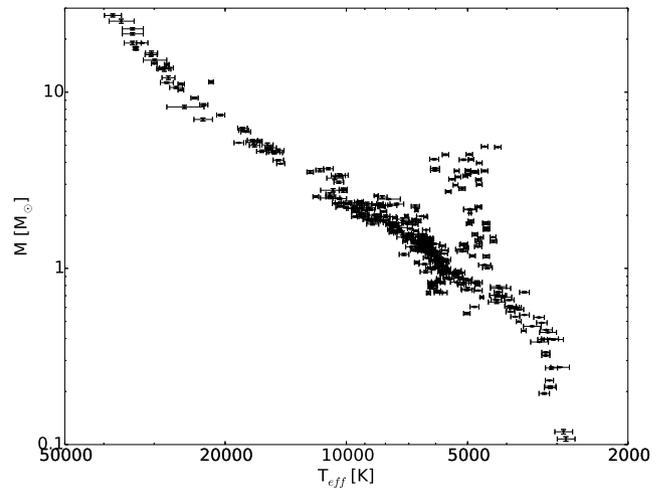}
 \caption{Mass versus effective temperature for the same objects as in Figure \ref{fig:fig5}.
}
 \label{fig:fig6}
\end{figure}

\begin{figure}
\centering
 \includegraphics[width=85mm,angle=0]{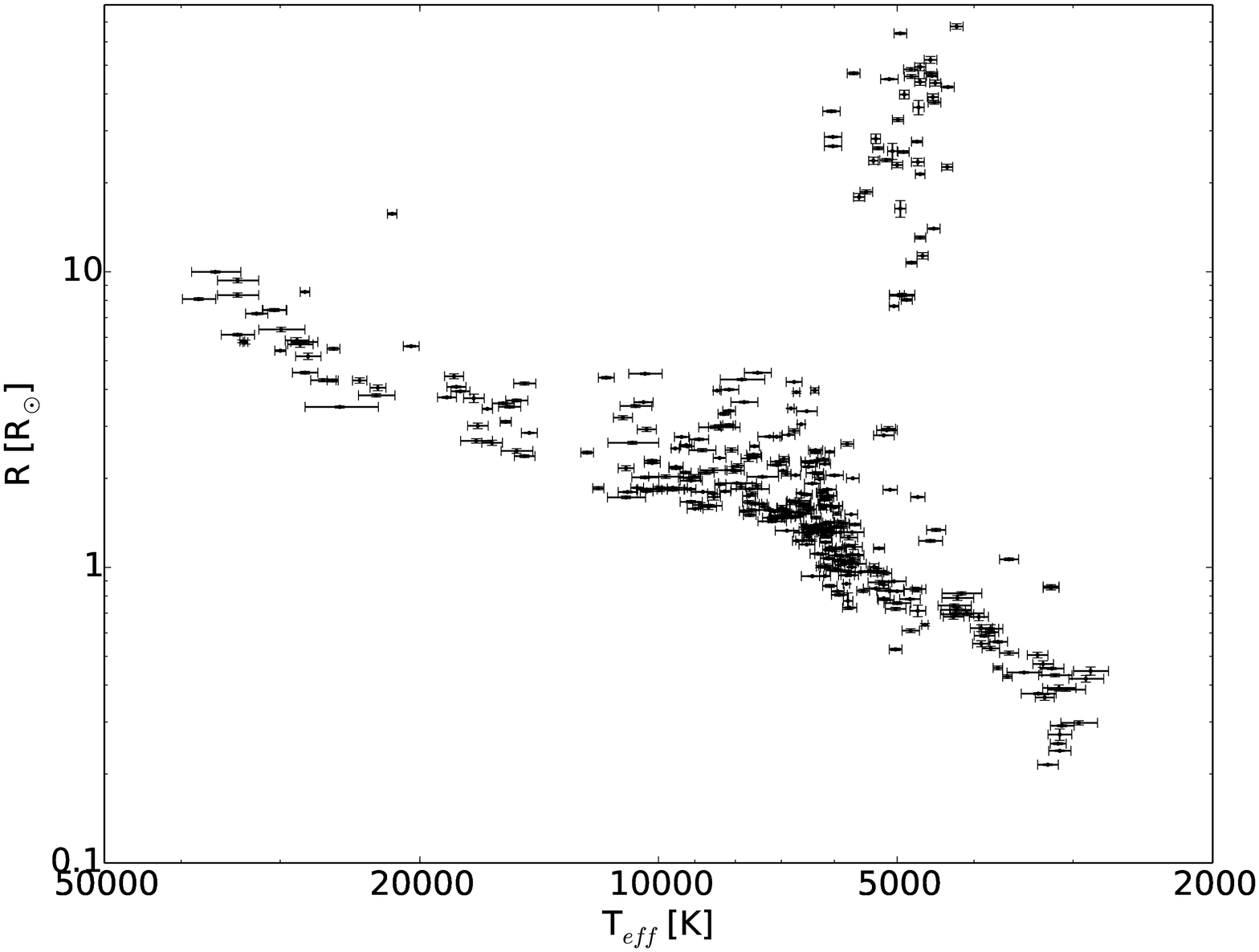}
 \caption{Radius versus effective temperature for the same objects as in Figure \ref{fig:fig5}.
}
 \label{fig:fig7}
\end{figure}

\begin{figure}
\centering
 \includegraphics[width=85mm,angle=0]{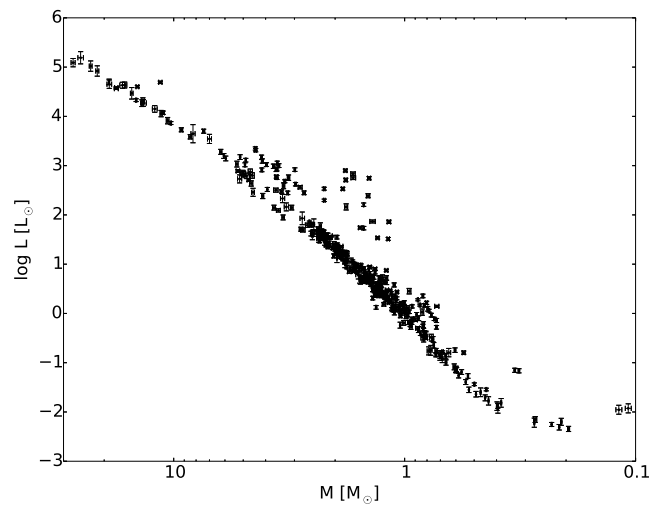}
 \caption{Luminosity versus mass for the same objects as in Figure \ref{fig:fig5}.
}
 \label{fig:fig8}
\end{figure}

\subsection{Effects of stellar evolution}
\label{stellarevolutioneffects}

Figures \ref{fig:fig6} and \ref{fig:fig7} display effective temperature versus mass and radius, respectively, 
for the full sample of DEB stars. As in Figure \ref{fig:fig5}, the effects of stellar evolution are evident. 
In the $M$ vs. $T_\text{eff}$ diagram stars move horizontally, towards cooler temperatures, as they evolve off the main sequence, assuming there is no mass loss.
In the $R$ vs. $T_\text{eff}$ diagram the effects are more evident, with displacements towards the upper right region, given the changes in both parameters. 

We updated the stellar luminosities for DEB stars, determining them from the corresponding effective temperatures and radii. 
Figure \ref{fig:fig8} shows a tight correlation between luminosity and mass on the main sequence. 
The scatter is mainly due to effects of stellar evolution and chemical abundance variations among the stars.

We present and discuss the effects of evolution for stars with metallicities in Section \ref{discussion} and Appendix \ref{appendix:evolution}.

\section{Methodology}
\label{method}

The inference of the stellar parameters from the {\sevensize PARSEC} v1.2S models was performed in a similar way to that described by \citet{jorgensen2005}. 
We obtained the different parameters from the likelihood function, ${\mathcal L}$, given in their Equation 4, 
and the prior probability density function, $f_\text{0}$, integrating them over the parameter space, namely $\tau$, $M_\text{0}$, and [Fe/H]. 
For instance, the estimated mean value and variance of the actual mass $M$ are respectively:

\begin{equation}
E(M)        \propto \iiint f_0~{\mathcal L}~M~dM_0~d\tau~d{\rm[Fe/H]}
\end{equation}
\begin{equation}
Var(M)      \propto \iiint f_0~{\mathcal L}~[M - E(M)]^2~dM_0~d\tau~d{\rm[Fe/H]}
\end{equation}

The constant of proportionality is chosen to ensure $\iiint f_\text{0}~{\mathcal L}~dM_\text{0}~d\tau~d$[Fe/H]=1.
We evaluated $f_\text{0}$ from the initial mass function of \citet{chabrier2001}, assuming flat priors on age and [Fe/H].

Means and variances for all parameters as well as a probability were calculated for every stellar evolutionary stage.  
But age was generally inferred from the posterior probability density function, as described in the next subsection.

\subsection{Posterior probability density functions}
We calculated the posterior probability density function $G(\tau)$ given by Equation 10 of \citet{jorgensen2005}, which is informative of stellar age $\tau$. 
We computed the median value (bisector of the area under the $G(\tau)$ function) and mode (global maximum of $G(\tau)$), 
as well as confirmed the mean value for age, obtained from Equation 1, after replacing $M$ by $\tau$.
The mean and median values of $G(\tau)$ can be severely biased, and the mode it is generally the best option to estimate ages \citep[][]{jorgensen2005}. 

Assuming the two components of the binary system formed at roughly the same time, it is possible to further constrain their age by multiplying 
the corresponding $G(\tau)$ functions to derive the mean, median, and mode of the combined function. 
If this combined profile is predominantly flat, we adopted the mean value of the averaged ages of the two binary components. 
This is the case of CM Dra.

We also obtained the posterior probability density function to yield the actual mass, although we found that it is generally much better constrained than the age.  
Equations 1 and 2 provide a fast, satisfactory calculation for the mass, as well as for any other stellar parameter apart from age. 
It is worth noting that for coeval DEB stars in clusters we could have combined their $G(\tau)$ functions in order to further constrain their age. 

\subsection{Most likely stellar evolutionary stages}
\label{probabilities}

We assessed a probability $Pr$ for every evolutionary stage, calculated from the area of the corresponding posterior probability density function. 
This allow us to distinguish between stellar evolutionary stages, with the sum of all probabilities being 100 per cent. 
We first chose that stage for which the probability is maximum, determining the corresponding stellar parameters. 
But we also inferred the parameters corresponding to the second most likely evolutionary stage when $Pr$<90 per cent for the most likely case. 
We then considered all the possible combinations of the selected evolutionary stages as explained in Section \ref{getparameters}.
Note $Pr$ values are just estimates, which partly depend on the grid of stellar evolutionary models used in the calculations.

In the following section we establish the accuracy and precision of {\sevensize PARSEC} v1.2S models under this Bayesian approach to predict the parameters of DEB stars. 
As example, we show the $G(\tau)$ functions for a system in Appendix \ref{appendix:likelihood}.

\section{Analysis and results}
\label{results}

\subsection{Inferred stellar parameters}
\label{getparameters}

Our strategy was to take on input three stellar parameters that can be empirically measured for the members of DEB systems to very high precision and accuracy,
namely $R$, $T_\text{eff}$, and [FeH]\footnote{The change in the photospheric metal mass fraction due to stellar evolution is assumed to be negligible.}, and their uncertainties, 
to infer other stellar parameters from the methodology described in Section \ref{method}. 
Especially, stellar mass, age, and distance, which can be compared with precise literature results.
We worked on Sample I (see Section \ref{selectionfeh}) and complemented our study using Sample II (see Section \ref{selectionnofeh}).

As input parameters, we could have adopted absolute magnitudes, e.g. $M_\text{V}$, or surface gravities. We chose to use radii because it is one of the fundamental 
parameters inferred from the light curves of eclipsing binaries, and as such it provides a solid reference for those systems, unlike $M_\text{V}$. Radius is also less directly 
connected to mass than $\log g$, and mass is the other parameter accurately known for DEB stars, and therefore using radius as input and mass as output reduces the 
correlations between their uncertainties. In addition, radius can be potentially derived with high accuracy combining angular diameters from (spectro)photometry and 
parallaxes (see Sect. \ref{inputs}). Effective temperatures were critically reviewed by \citet{torres2010} and we have found an excellent agreement for the 
84 systems in common with DEBCat (see Sect. \ref{comparingDEBcat}). Regarding [Fe/H], our tests show that output parameters for nearby stars do not change 
much if assuming solar metallicity (see Sect. \ref{getmass}).

\begin{figure}
\centering
 \includegraphics[width=85mm,angle=0]{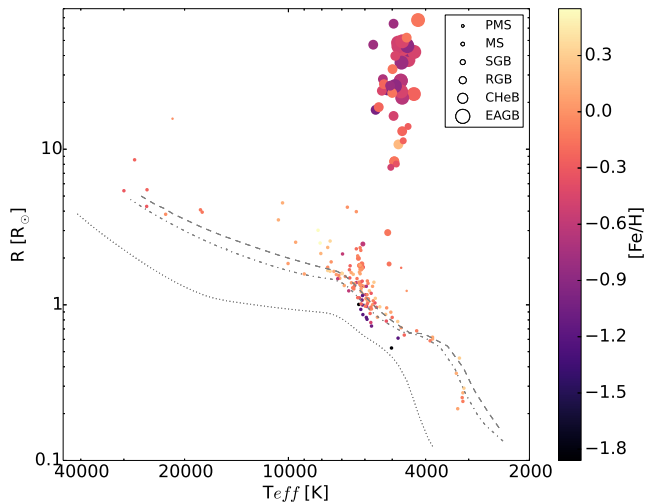}
 \caption{Measured radius versus effective temperature for stars in Sample I, colour-coded by metallicity and size-coded by stellar evolutionary stage. 
 Dotted, dash-dotted and dashed lines correspond to the ZAMS for [Fe/H]=-2.1, 0.0 and 0.4, respectively.
}
 \label{fig:fig9}
\end{figure}

Tables \ref{tab:sampleI} and \ref{tab:sampleII} list\footnote{The full list of DEB stars is available in the online version of \mnras.} the literature stellar parameters for the binary components of Sample I and Sample II, respectively. 
These parameters are radius ($R$), effective temperature ($T_\text{eff}$), iron-to-hydrogen ratio [Fe/H], mass ($M$), surface gravity ($g$), luminosity ($L$), 
spectral type ($SpT$)/luminosity class ($LC$), colour excess $E(B-V)$, and distance ($d$). %, and period ($P$). 
Well constrained ages for 25 systems in clusters are given and compared with our results in Section \ref{getage}.
We provide references for the tabulated $E(B-V)$ and distances in Section \ref{getdistance}. 
Tables \ref{tab:sampleI} and \ref{tab:sampleII} contain two rows for every system, which respectively correspond to the primary and secondary. 
Systems are sorted by mass of the primary. 
Note [Fe/H] for SW CMa is out of the grid of theoretical models, although some of them are within uncertainty limits.
Therefore, the inferred parameters for the two binary components of SW CMa must be carefully taken.

Tables \ref{tab:sampleIb} and \ref{tab:sampleIIb} present\footnote{See previous footnote.} the inferred parameters for the binary components of Sample I and Sample II, respectively. 
These are absolute magnitudes in the B-band ($M_\text{B}$) and the V-band ($M_\text{V}$), bolometric correction ($BC \equiv M_{bol}$-$M_\text{V}$), mass ($M_\text{m}$), evolutionary stage ($ES$), estimated probability ($Pr$), as well as the distance ($d_\text{m}$) and age ($\tau$) for every system. 
We also indicate the grid used for the inference of these parameters ($O$: original grid; $I$: interpolated grid). 
We have introduced a criterion of probability to present the most likely stellar parameters.
When the two binary components present an inferred evolutionary stage with $Pr>$90 per cent, 
we only provide a single combined age and a distance. 
If $Pr<$90 per cent for one of the two components of the system we include the second most likely stage for such component, 
yielding two pairs for which we compute the combined ages and distances. 
When $Pr<$90 per cent for the two components there are four possible combinations, every one with their corresponding ages and distances. 
In summary, Table \ref{tab:sampleIb} and \ref{tab:sampleIIb} contain either two, four or eight rows for every system, with the parameters of the primary and secondary. 
The different number of rows is due to the case given by the criterion of probability.
Binaries are sorted in the same way as in Tables \ref{tab:sampleI} and \ref{tab:sampleII}. 

Figure \ref{fig:fig9} displays the $R$ vs. $T_\text{eff}$ diagram, colour-coded by [Fe/H], for Sample I. 
The three input parameters are separated into groups corresponding to the most likely stellar evolutionary stage. 
Evolution effects are evident. Note the loci of low metallicity stars and the most evolved stars. 
In Appendix \ref{appendix:zams} we provide polynomial fittings to the theoretical zero-age main-sequence at [Fe/H]=-2.1, 0.0, and +0.4 dex.

\begin{landscape}
 \begin{table}
  \caption{
Literature stellar parameters of DEB stars in Sample I, with known metallicities: radius ($R$), effective temperature ($T_\text{eff}$), 
iron abundance [Fe/H], and mass ($M$), surface gravity ($\log g$), luminosity ($\log L$), spectral type ($SpT$)/luminosity class ($LC$), 
colour excess $E(B-V)$, and distance $d$. 
References for the literature values are given in the main text. A full version of this table is available online.
}
  \label{tab:sampleI}
  \begin{tabular}{ccccccccccc}
    \hline
    System & $R$               & $T_\text{eff}$ & [Fe/H]      & $M$              & $\log g$              &  $\log L$                 &  $SpT/LC$  & $E(B-V)$        &  $d$             \\
                 & (R$_{\sun}$) & ($\,K$)     & (dex)   &  (M$_{\sun}$) & ($g$: cm s$^{-2}$)  & ($L$:  L$_{\sun}$)     &                    & (mag)         &  (\pcw)      \\
    \hline
CM Dra & 0.2534 $\pm$ 0.0019 & 3133 $\pm$ 72 & -0.30 $\pm$ 0.12 & 0.2310 $\pm$ 0.0009 & 4.994 $\pm$ 0.007 & -2.25 $\pm$ 0.04 & M4.5V & 0 & 14.850 $\pm$ 0.011 \\
                        & 0.2396 $\pm$ 0.0015 & 3119 $\pm$ 101 & $\shortparallel$ & 0.2141 $\pm$ 0.0009 & 5.009 $\pm$ 0.006 & -2.31 $\pm$ 0.06 & M4.5V & $\shortparallel$ & $\shortparallel$ \\
PTFEB 132.707+19.810 & 0.363 $\pm$ 0.008 & 3258 $\pm$ 90 & 0.14 $\pm$ 0.04 & 0.3953 $\pm$ 0.0020 & 4.915 $\pm$ 0.019 & -1.87 $\pm$ 0.05 & M3.5V & 0 & 187 $\pm$ 3 \\
                        & 0.272 $\pm$ 0.012 & 3119 $\pm$ 108 & $\shortparallel$ & 0.2098 $\pm$ 0.0014 & 4.89 $\pm$ 0.04 & -2.20 $\pm$ 0.07 & M4.3V & $\shortparallel$ & $\shortparallel$ \\
\hline
  \end{tabular}
 \end{table}
\end{landscape}

\begin{landscape}
 \begin{table}
  \caption{
Literature stellar parameters of DEB stars in Sample II, with adopted solar metallicity: radius ($R$), effective temperature ($T_\text{eff}$), 
iron abundance [Fe/H], and mass ($M$), surface gravity ($\log g$), luminosity ($\log L$), spectral type ($SpT$)/luminosity class ($LC$), 
colour excess $E(B-V)$, and distance $d$.
References for the literature values are given in the main text. A full version of this table is available online.
}
  \label{tab:sampleII}
  \begin{tabular}{ccccccccccc}
    \hline
    System & $R$               & $T_\text{eff}$ & [Fe/H]      & $M$              & $\log g$              &  $\log L$                 &  $SpT/LC$  & $E(B-V)$        &  $d$    \\
                 & (R$_{\sun}$) & ($\,K$)     & (dex)   &  (M$_{\sun}$) & ($g$: cm s$^{-2}$)  & ($L$:  L$_{\sun}$)     &                    & (mag)         &  (\pcw) \\
    \hline
EPIC 203710387 & 0.446 $\pm$ 0.014 & 2851 $\pm$ 144 & 0.00 $\pm$ 0.19 & 0.108 $\pm$ 0.003 & 4.170 $\pm$ 0.022 & -1.93 $\pm$ 0.09 & M5V & 0 &  \\
                        & 0.420 $\pm$ 0.011 & 2891 $\pm$ 146 & $\shortparallel$ & 0.118 $\pm$ 0.004 & 4.263 $\pm$ 0.024 & -1.95 $\pm$ 0.09 & M4.5V & $\shortparallel$ &  \\
UScoCTIO 5 & 0.862 $\pm$ 0.012 & 3199 $\pm$ 74 & 0.00 $\pm$ 0.19 & 0.3336 $\pm$ 0.0022 & 4.090 $\pm$ 0.012 & -1.15 $\pm$ 0.04 & M4.5V & 0.246 $\pm$ 0.004 &  \\
                        & 0.852 $\pm$ 0.013 & 3199 $\pm$ 74 & $\shortparallel$ & 0.3200 $\pm$ 0.0022 & 4.082 $\pm$ 0.012 & -1.16 $\pm$ 0.04 & M4.5V & $\shortparallel$ &  \\
\hline
  \end{tabular}
 \end{table}
\end{landscape}

\begin{landscape}
 \begin{table}
  \caption{
Inferred stellar parameters of DEB stars in Sample I: absolute magnitudes in the B-band ($M_\text{B}$) and the V-band ($M_\text{V}$), bolometric correction ($BC$), mass ($M_\text{m}$), evolution stage ($ES$), probability $Pr$, distance ($d_\text{m}$), age ($\tau$), and grid used. A full version of this table is available online.}
  \label{tab:sampleIb}
  \begin{tabular}{cccccccccc}
    \hline
    Binary & $M_\text{B}$       &  $M_\text{V}$         & $BC$        &    $M_\text{m}$          & $ES$  &  $Pr$    & $d_\text{m}$      &  $\tau$  & Grid \\
                 & (mag)     & (mag)        &  (mag)  &    (M$_{\sun}$)  &            &              & (\pcw)      & (Gyr)  &        \\
    \hline
CM Dra A & 14.47 $\pm$ 0.26 & 12.88 $\pm$ 0.26 & -2.50 $\pm$ 0.18 & 0.235 $\pm$ 0.004 & MS & 94 & 13.7 $\pm$ 2.8 & 7 $_{- 4 }^{+ 4 }$ & $I$ \\
CM Dra B & 14.7 $\pm$ 0.3 & 13.1 $\pm$ 0.3 & -2.59 $\pm$ 0.23 & 0.219 $\pm$ 0.003 & MS & 95 & $\shortparallel$ & 7 $_{- 4 }^{+ 4 }$ & $I$ \\
PTFEB 132.707+19.810 A & 13.43 $\pm$ 0.14 & 11.95 $\pm$ 0.14 & -2.43 $\pm$ 0.08 & 0.359 $\pm$ 0.010 & MS & 80 & 205 $\pm$ 69 & 3.2 $_{- 0.5 }^{+ 7 }$ & $I$ \\
PTFEB 132.707+19.810 B & 14.81 $\pm$ 0.26 & 13.27 $\pm$ 0.26 & -2.91 $\pm$ 0.13 & 0.254 $\pm$ 0.014 & MS & 96 & $\shortparallel$ & $\shortparallel$ & $I$ \\
PTFEB 132.707+19.810 A & 13.37 $\pm$ 0.14 & 11.89 $\pm$ 0.14 & -2.39 $\pm$ 0.08 & 0.367 $\pm$ 0.011 & PMS & 20 & 210 $\pm$ 71 & 0.77 $_{- 0.09 }^{+ 1.2 }$ & $I$ \\
PTFEB 132.707+19.810 B & 14.81 $\pm$ 0.26 & 13.27 $\pm$ 0.26 & -2.91 $\pm$ 0.13 & 0.254 $\pm$ 0.014 & MS & 96 & $\shortparallel$ & $\shortparallel$ & $I$ \\
\hline
  \end{tabular}
 \end{table}
\end{landscape}

\begin{landscape}
 \begin{table}
  \caption{
Inferred stellar parameters of DEB stars in Sample II: absolute magnitudes in the B-band ($M_\text{B}$) and the V-band ($M_\text{V}$), bolometric correction ($BC$), mass ($M_\text{m}$), evolution stage ($ES$), probability $Pr$, distance ($d_\text{m}$), age ($\tau$), and grid used. A full version of this table is available online.}
  \label{tab:sampleIIb}
  \begin{tabular}{cccccccccc}
    \hline
    Binary & $M_\text{B}$       &  $M_\text{V}$         & $BC$        &    $M_\text{m}$          & $ES$  &  $Pr$    & $d_\text{m}$      &  $\tau$  & Grid \\
                 & (mag)     & (mag)        &  (mag)  &    (M$_{\sun}$)  &            &              & (\pcw)      & (Gyr)  &        \\
    \hline
EPIC 203710387 A & 14.3 $\pm$ 0.8 & 12.8 $\pm$ 0.8 & -3.3 $\pm$ 0.6 & 0.22 $\pm$ 0.08 & PMS & 70 & 22 $\pm$ 8 & 1.94 $_{- 0.27 }^{+ 0.26 }$ & $I$ \\
EPIC 203710387 B & 12.97 $\pm$ 0.27 & 11.55 $\pm$ 0.29 & -2.37 $\pm$ 0.18 & 0.415 $\pm$ 0.013 & MS & 63 & $\shortparallel$ & $\shortparallel$ & $I$ \\
EPIC 203710387 A & 14.3 $\pm$ 0.8 & 12.8 $\pm$ 0.8 & -3.3 $\pm$ 0.6 & 0.22 $\pm$ 0.08 & PMS & 70 & 25 $\pm$ 12 & 0.034 $_{- 0.009 }^{+ 0.018 }$ & $I$ \\
EPIC 203710387 B & 14.0 $\pm$ 0.9 & 12.6 $\pm$ 0.8 & -3.0 $\pm$ 0.6 & 0.26 $\pm$ 0.10 & PMS & 37 & $\shortparallel$ & $\shortparallel$ & $I$ \\
EPIC 203710387 A & 12.75 $\pm$ 0.26 & 11.34 $\pm$ 0.27 & -2.32 $\pm$ 0.16 & 0.438 $\pm$ 0.015 & MS & 30 & 42 $\pm$ 11 & 13.0 $_{- 7 }^{+ 0.8 }$ & $I$ \\
EPIC 203710387 B & 12.97 $\pm$ 0.27 & 11.55 $\pm$ 0.29 & -2.37 $\pm$ 0.18 & 0.415 $\pm$ 0.013 & MS & 63 & $\shortparallel$ & $\shortparallel$ & $I$ \\
EPIC 203710387 A & 12.75 $\pm$ 0.26 & 11.34 $\pm$ 0.27 & -2.32 $\pm$ 0.16 & 0.438 $\pm$ 0.015 & MS & 30 & 50 $\pm$ 24 & 1.94 $_{- 0.3 }^{+ 0.28 }$ & $I$ \\
EPIC 203710387 B & 14.0 $\pm$ 0.9 & 12.6 $\pm$ 0.8 & -3.0 $\pm$ 0.6 & 0.26 $\pm$ 0.10 & PMS & 37 & $\shortparallel$ & $\shortparallel$ & $I$ \\
UScoCTIO 5 A & 11.43 $\pm$ 0.27 & 10.04 $\pm$ 0.28 & -2.41 $\pm$ 0.19 & 0.41 $\pm$ 0.07 & PMS & 100 & 163 $\pm$ 38 & 0.0116 $_{- 0.0026 }^{+ 0.0029 }$ & $I$ \\
UScoCTIO 5 B & 11.46 $\pm$ 0.27 & 10.06 $\pm$ 0.28 & -2.41 $\pm$ 0.19 & 0.41 $\pm$ 0.07 & PMS & 100 & $\shortparallel$ & $\shortparallel$ & $I$ \\
    \hline
  \end{tabular}
 \end{table}
\end{landscape}

\subsection{Stellar mass comparison}
\label{getmass}

\begin{figure}
\centering
 \includegraphics[width=85mm,angle=0]{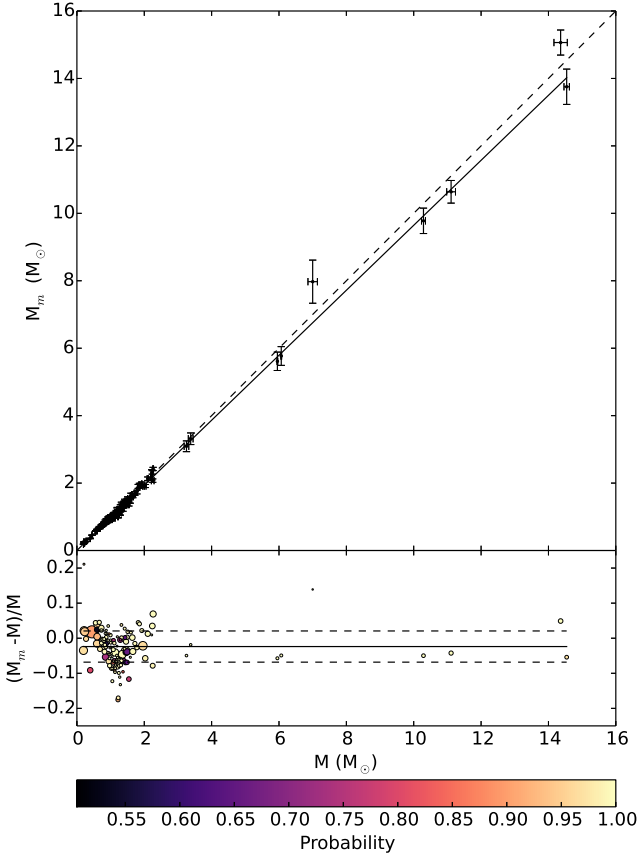}
 \caption{Upper panel plots modelled mass $M_\text{m}$ versus dynamical mass $M$ for stars in Sample I that are most likely in the main sequence. 
 Continuous and dashed lines respectively stand for best linear fit and 1:1 relation.
 Lower panel shows $\frac{M_\text{m}-M}{M}$ against $M$ for these stars. 
 Symbols are colour-coded by probability and size-scaled by the square of inverse uncertainty of $\frac{M_\text{m}-M}{M}$, normalised to the maximum value. 
 Continuous and dashed lines respectively stand for weighted mean and weighted standard deviation limits.
 }
 \label{fig:fig10}
\end{figure}

\begin{figure}
\centering
 \includegraphics[width=85mm,angle=0]{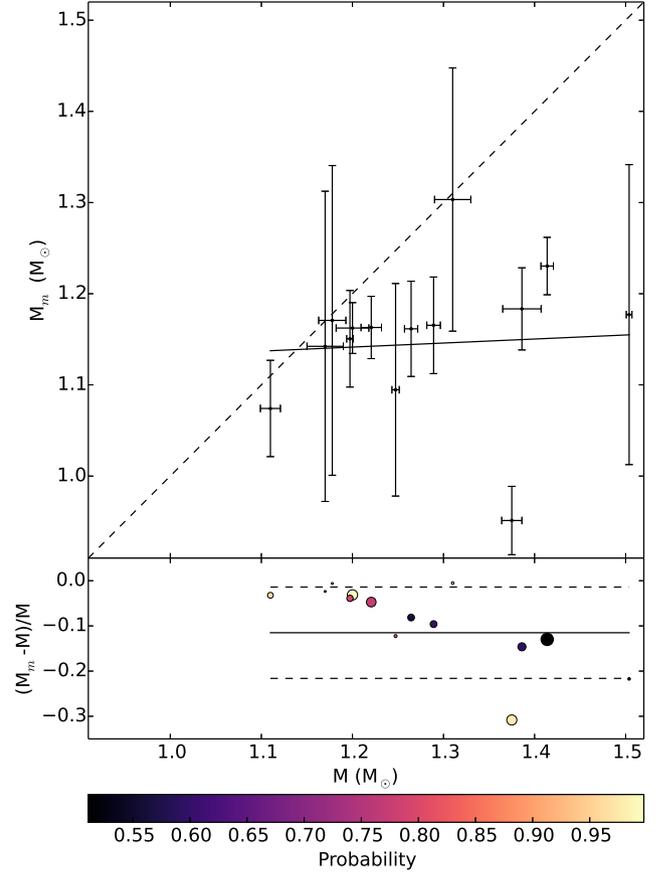}
 \caption{Upper panel displays modelled mass $M_\text{m}$ versus dynamical mass $M$ for stars in Sample I that are most likely in the SGB and RGB stages. 
 Lower panel shows $\frac{M_\text{m}-M}{M}$ against $M$ for these stars. 
 Lines and symbols as in Figure \ref{fig:fig10}. 
 }
 \label{fig:fig11}
\end{figure}

In order to assess the accuracy and precision of the {\sevensize PARSEC} v1.2S models in combination with the Bayesian method described in Section \ref{method}, 
we compare the accurate dynamical masses of the selected DEB stars with their respective inferred values. 

We split up Sample I into four groups according with their most likely stellar evolutionary stages $ES$.
We arrive at 129, 14, 19, and 17 binaries that are plausibly on the MS, SGB or RGB (hereafter, S/RGB), in the CHeB phase, and in the EAGB stage, respectively.
If adding the objects from Sample II, the statistics increase for MS and S/RGB to 254 and 19 objects, respectively.
We could not provide a reliable solution for the few objects which are most likely PMS. 

\begin{figure}
\centering
 \includegraphics[width=85mm,angle=0]{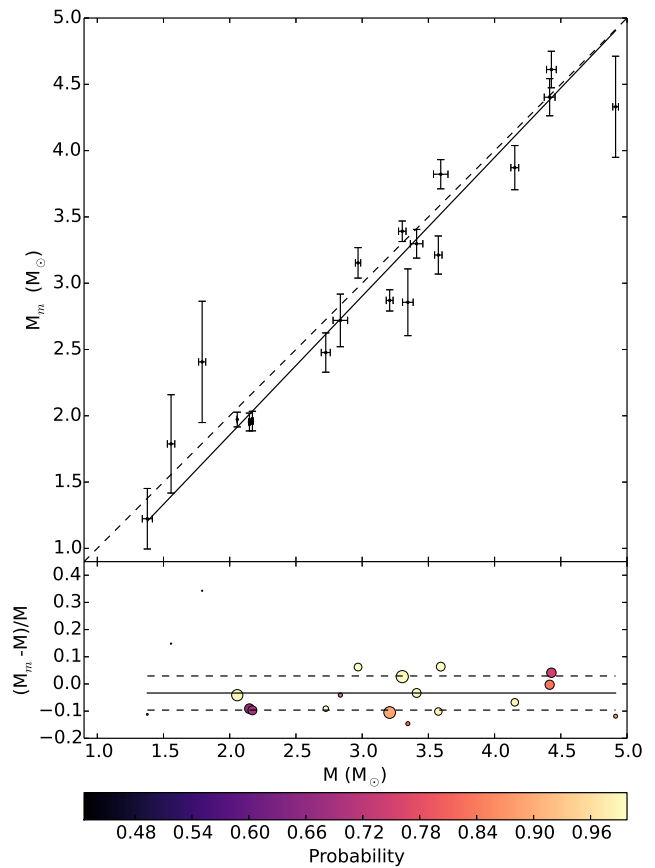}
 \caption{Upper panel plots modelled mass $M_\text{m}$ versus dynamical mass $M$ for stars in Sample I that are most likely in the CHeB phase. 
 Lower panel shows $\frac{M_\text{m}-M}{M}$ against $M$ for these stars. 
 Lines and symbols as in Figure \ref{fig:fig10}. 
 }
 \label{fig:fig12}
\end{figure}

\begin{figure}
\centering
 \includegraphics[width=85mm,angle=0]{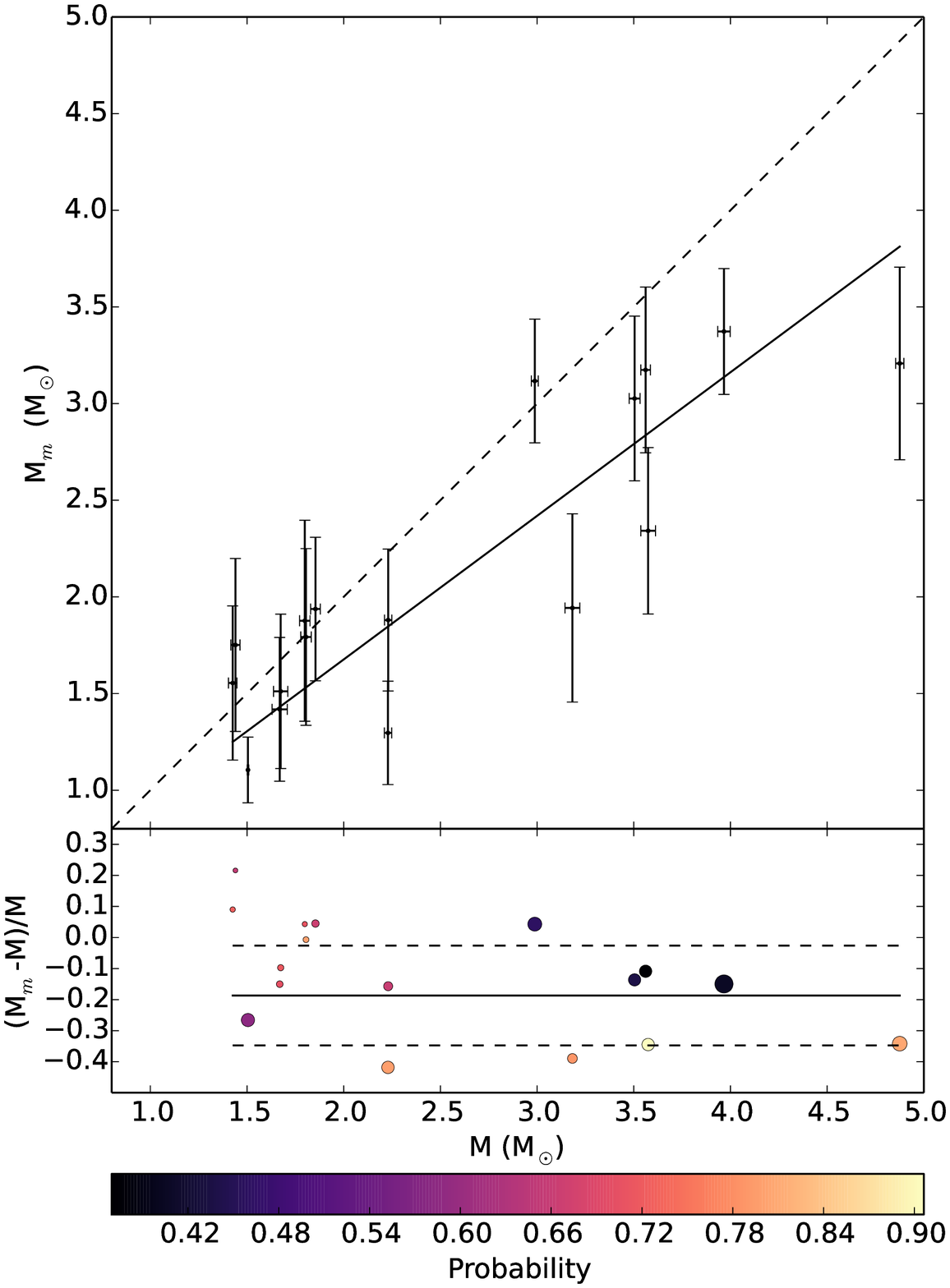}
 \caption{Upper panel displays modelled mass $M_\text{m}$ versus dynamical mass $M$ for stars in Sample I that are most likely in the EAGB stage. 
 Lower panel shows $\frac{M_\text{m}-M}{M}$ against $M$ for these stars. 
 Lines and symbols as in Figure \ref{fig:fig10}. 
 }
 \label{fig:fig13}
\end{figure}

The upper panels of Figures \ref{fig:fig10}--\ref{fig:fig13} show $M_\text{m}$ vs. $M$ and the best linear fit obtained from 
a weighted orthogonal distance regression procedure \citep[][]{boggs1992} for every group of Sample I.
The algorithm takes into account the uncertainties in $M_\text{m}$ and $M$ to compute the slope and offset of the linear regression.
The lower panels of Figures \ref{fig:fig10}--\ref{fig:fig13} show the relative residuals $\frac{M_\text{m}-M}{M}$ against $M$ for every group.
We obtained the weighted mean and weighted standard deviation of the aforementioned relative residuals for each group. 
We first used as weights the inverse of the squared uncertainties, and then the product of these with the assessed probabilities $Pr$.
Both approaches produce similar results. 
Pearson correlation coefficients weighted by the inverse of the squared uncertainties are also calculated.
Table \ref{tab:groupeddeviations} summarises our results.

For the four groups, values of weighted mean $\frac{M_\text{m}-M}{M}$ indicate modelled masses are on average lower than dynamical masses. 
Although they are comparable to the standard deviations, they are significantly greater than the precision of the dynamical masses for S/RGB and EAGB stars. 
It is worth noting that residuals are greater when using the actual mass derived with the contributions of all stellar evolutionary stages instead 
of selecting that of the most likely stage.
The values of $\frac{M_\text{m}-M}{M}$ can be interpreted in terms of the computed slopes and offsets. 
For example, the MS group presents a very small offset and a slope slightly lower than one while CHeB stars present a slope above one but a negative offset. 
Note that the result for MS stars is consistent employing Sample I and the combination of Sample I and Sample II.
We also made a fitting for stars closer than 300\pc in Sample I assuming solar metallicity, to gauge the impact of lacking metallicity information.
The results are consistent with each other and support they are statistically meaningful.

A recent work by \citet{ghezzi2015} has concluded that the masses of evolved stars inferred from {\sevensize PARSEC} stellar evolutionary models are not significantly 
affected by systematic errors. These authors included 26 binaries, 23 of them in common with our Sample I. 
We found that most of the literature values for the dynamical masses are the same, 
which is confirmed by a mean averaged relative difference of -0.004$\pm$0.005 and a Pearson correlation coefficient $r$=0.999995.
Given this consistency, we applied the weighted orthogonal distance regression procedure of \citet{boggs1992} to the 26 binaries of \citet{ghezzi2015}
and to all the 41 RGB, CHeB, and EAGB binaries in Sample I\footnote{The number of objects increases by only one if including Sample II.} (see Table \ref {tab:comparetoGhezzi}). 
Note the excellent agreement between the two sets of results. 
It is relevant to highlight, however, that our separation in different evolutionary stages is key to show the prominent discrepancies for the S/RGB and EAGB groups of evolved stars.

In summary, our analysis yields good agreement between theoretical and dynamical masses for stars on the MS and CHeB phases, 
while predictions for sub-giants and giants as well as for stars on the EAGB stage are significantly discrepant from dynamical values. 
It is important to utilise the right metallicity for every star, although one can assume solar metallicity for nearby objects. 
The use of Sample II allows us to complement our tests aimed at determining the accuracy and precision of the {\sevensize PARSEC} v1.2S library.

\begin{table*}
\caption{Number of points N, mean $\pm$ standard deviation of $\frac{M_\text{m}-M}{M}$ weighted by $\sigma^{-2}$ ($w_\text{1}$) and by normalised probability $\times~\sigma^{-2}$ ($w_\text{2}$), Pearson correlation coefficient weighted by $w_\text{1}$, slope and offset computed using a weighted orthogonal distance regression procedure \citep[][]{boggs1992}, and mass range for different groups of DEB stars in Sample I and Sample I + II.}
  \label{tab:groupeddeviations}
  \begin{tabular}{cccccccccc}
    \hline
    Group &   N & $\langle\frac{M_\text{m}-M}{M}\rangle_{\text{w}_\text{1}}$  & $\langle\frac{M_\text{m}-M}{M}\rangle_{\text{w}_\text{2}}$   & $r_{\text{w}_\text{1}}$ & slope  & offset       & mass range    \\
               &      &                                                  &                                               &                                                       &                                 & (M$_\odot$)   & (M$_\odot$)     \\
    \hline
MS (I)                   & 129 & -0.02$\pm$0.04 & -0.02$\pm$0.04 &  0.998  & 0.964$\pm$0.006 & 0.010$\pm$0.003 & 0.2 -- 14.5 \\
MS (I$+$II)               & 254 & -0.02$\pm$0.06 & -0.02$\pm$0.05 & 0.998  &  0.955$\pm$0.007 & 0.016$\pm$0.004 & 0.1 -- 14.5 \\
S/RGB (I)              & 14 & -0.12$\pm$0.09 & -0.12$\pm$0.10 &  0.060  & 0.04$\pm$0.24 & 1.1$\pm$0.3     & 1.1 -- 1.5 \\
S/RGB (I$+$II)          & 19 & -0.11$\pm$0.09 & -0.11$\pm$0.10 &  0.133  & 0.12$\pm$0.20 & 0.99$\pm$0.25 & 1.1 -- 1.5 \\
CHeB (I)               & 19 & -0.03$\pm$0.06 & -0.03$\pm$0.06 &  0.973 & 1.05$\pm$0.05 & -0.24$\pm$0.16 & 1.4 -- 4.9 \\
EAGB (I)              & 17 & -0.19$\pm$0.15 & -0.21$\pm$0.16 &  0.856 & 0.74$\pm$0.10 & 0.19$\pm$0.26 & 1.4 -- 4.9 \\
  \hline
  \end{tabular}
\end{table*}

\begin{table*}
\caption{Number of points N, mean $\pm$ standard deviation of $\frac{\Delta M}{M}$ weighted by $\sigma^{-2}$ ($w_\text{1}$) and by normalised probability $\times~\sigma^{-2}$ ($w_\text{2}$), Pearson correlation coefficient weighted by $w_\text{1}$, slope and offset computed using a weighted orthogonal distance regression procedure \citep[][]{boggs1992}, and mass range for different groups of DEB stars: 
1) the full sample of 26 giants of \citet{ghezzi2015}, with $\Delta M$ being the relative difference between their modelled masses with respect to the dynamical masses; and
2) all the 41 giants (i.e., stars with inferred evolutionary stages RGB, CHeB, or EAGB) in Sample I.}
  \label{tab:comparetoGhezzi}
  \begin{tabular}{ccccccccc}
    \hline
    N & $\langle\frac{M_\text{m}-M}{M}\rangle_{\text{w}_\text{1}}$  & $\langle\frac{M_\text{m}-M}{M}\rangle_{\text{w}_\text{2}}$   & $r_{\text{w}_\text{1}}$ & slope  & offset       & mass range    \\
       &                                                  &                                               &                                                       &                                 & (M$_\odot$)   & (M$_\odot$)     \\
    \hline
    26 & -0.01$\pm$0.06 & --                       &  0.964   &  1.04$\pm$0.05   & -0.12$\pm$0.13    & 1.2 -- 4.4 \\
    41 & -0.04$\pm$0.08 & -0.04$\pm$0.08 & 0.956   & 1.01$\pm$0.04    & -0.16$\pm$0.11    &  1.2 -- 4.9 \\
  \hline
  \end{tabular}
\end{table*}

\subsection{Stellar age comparison}
\label{getage}

Age can be hard to constrain, particularly for low-mass field dwarfs. 
Inferred ages can be significantly dissimilar for the different combinations resulting from the criterion of probability, as shown for EPIC 203710387 in Table \ref{tab:sampleIIb}. 
We conclude a posteriori that this system likely consists of two PMS components despite the highest $Pr$ for its secondary corresponds to a MS star. 
The inferred mass for the secondary is significantly lower if this is on the PMS phase (0.26 $\pm$ 0.10 $M_{\odot}$) than if it is on the MS (0.415 $\pm$ 0.013 $M_{\odot}$). 
The discrepancy with the dynamical mass (0.118 $\pm$ 0.004 $M_{\odot}$) is quite significant anyway. The primary also shows a significant discrepancy. 
Note PMS phase is particularly hard to model and we assumed the system has solar metallicity. 
EPIC 203710387 is one of the four systems in Sample II for which we compiled ages available in the literature. 
We could do the same for 21 systems in Sample I. 

\begin{figure}
\centering
 \includegraphics[width=85mm,angle=0]{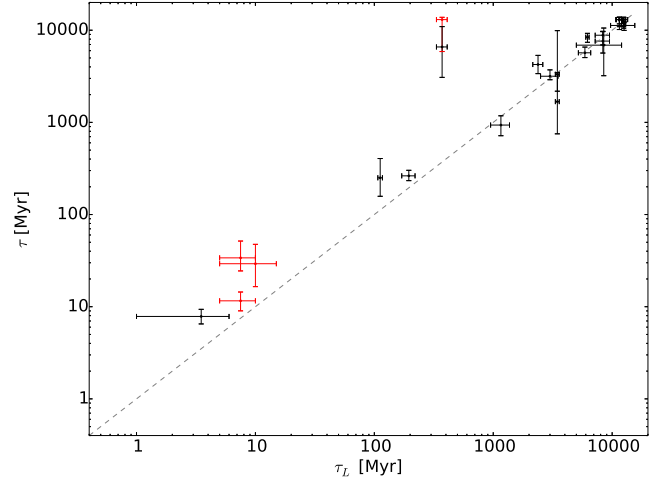}
 \caption{Ages inferred from this work against those found in the literature for binaries for 21 systems
 in Sample I (black dots) and four systems in Sample II (red dots). 
 The two outliers are CU Cnc and YY Gem, for which our ages are significantly higher.
}
 \label{fig:fig14}
\end{figure}

In most cases, literature ages come from membership of the binary in a cluster, from the turn-off location, 
but in some cases they are derived from the white dwarf cooling sequence, the analysis of hot star populations, or asteroseismology.
Both for globular and galactic clusters there is good agreement between the published ages and those inferred in this work, 
as illustrated in Fig. \ref{fig:fig14}.

M55 V54 belongs to the globular cluster M55 \citep[12.5$\pm$1.0 Gyr,][]{jimenez1998}. 
There are two binaries in NGC 6362 \citep[12.5 $\pm$ 0.5 Gyr,][]{kaluzny2015}, three members of M4 \citep[11.6 $\pm$ 0.6 Gyr,][]{bedin2009}, 
and one in 47 Tuc \citep[9.7 -- 15.5 Gyr,][]{salaris2002, zoccali2001}.
Some of the eclipsing binaries are in younger, open clusters, such as V565 Lyr and V568 Lyr in NGC 6791 \citep[7.2--9.5 Gyr,][]{grundahl2008},
V785 Cep in NGC 188 \citep[6.2 $\pm$ 0.2 Gyr,][]{meibom2009}, HAT-TR 318-007 and WOCS 12009 in NGC 2682 (M 67) \citep[3.46$\pm$0.13 Gyr,][]{sandquist2018},
V375 Cep in NGC 7142 \citep[3.0 $\pm$ 0.5 Gyr,][]{straizys2014}, WOCS 40007 in NGC 6819 \citep[2.38 $\pm$ 0.23 Gyr,][]{brewer2016}, 
KIC 9777062 in NGC 6811 \citep[0.95--1.37 Gyr][]{sandquist2016}, V906 Sco in M7 (NGC 6475) \citep[170--220 Myr,][]{sestito2003}, 
and V1229 Tau in the Pleiades \citep[112 $\pm$ 5 Myr,][]{dahm2015}.
The youngest stars for which we have reliable ages are V760 Sco in the Sco OB 2-2 association \citep[5--15 Myr;][]{feiden2016}, 
EPIC 203710387 and V Sco CTIO5 in the Upper Sco OB association \citep[5--10 Myr;][]{rizzuto2015}, 
and V578 Mon in the cluster NGC 2244 \citep[1--6 Myr,][]{bonatto2009}. 

We also include estimates for the age of two well-known binary systems. 
\citet{mamajek2008} respectively derived ages of 6.6 and 5.2 Gyr for the main and secondary components of $\alpha$ Cen A using activity-rotation diagnostics. 
In addition, we found a consistent value with the estimated age of 8.5$\pm$3.5 Gyr for the low-mass DEB binary CM Dra, 
which was derived from an age estimate of its common proper motion companion WD 1633+572 \citep[][]{feiden2014}.

The most discrepant cases are CU Cnc and YY Gem. 
YY Gem is physically associated with Castor A/B binary system and CU Cnc is likely a member of the same moving group, 
dated from the Castor A/B at $\sim$ 370 Myr \citep[][]{torres2002,ribas2003}.  Given our success with other systems containing components with similar masses, 
our results speak against a common age for these systems and the Castor A/B binary, but we refer the reader to the aforementioned papers, 
where significant discrepancies between the parameters for the stars in these systems and stellar evolutionary models have been reported. 
Note that {\sevensize PARSEC} v1.2S library introduces significant improvements for low mass dwarfs.

\subsection{Confusion diagrams}
\label{confusion}

Statistical parameter inference is subject to confusion because multiple distinct evolutionary paths pass through the volume defined by the resolution element, 
and therefore the resulting estimates may be uncertain and biased. 
\citet{delburgo2016} ascertained the parameters of planet host HD 209458 after calibrating the predictions from {\sevensize PARSEC} models against the Sun, 
given the similarity between the two stars. Unfortunately, we do not know so well other stars across the parameter space.

To get a general picture of the impact of confusion we performed simulations by using a collection of stellar evolution models distributed over the $M_\text{V}$ vs. $B-V$ diagram. 
We adopted $\sigma$([Fe/H])=0.10 dex, $\sigma(R)/R$=0.009, and $\sigma(T_\text{eff})/T_\text{eff}$=0.018, 
which are representative of Sample I. A total of $N$= 469,941 distinct simulations were accomplished.
We compare the input parameters with the mean values and variances obtained from Equations 1 and 2 in Section \ref{method}. 
We map the relative residuals in mean mass and mean age. 
But for these simulations we averaged the parameters without distinguishing among stellar evolutionary stages.

\begin{figure*}
\centering
 \includegraphics[width=170mm,angle=0]{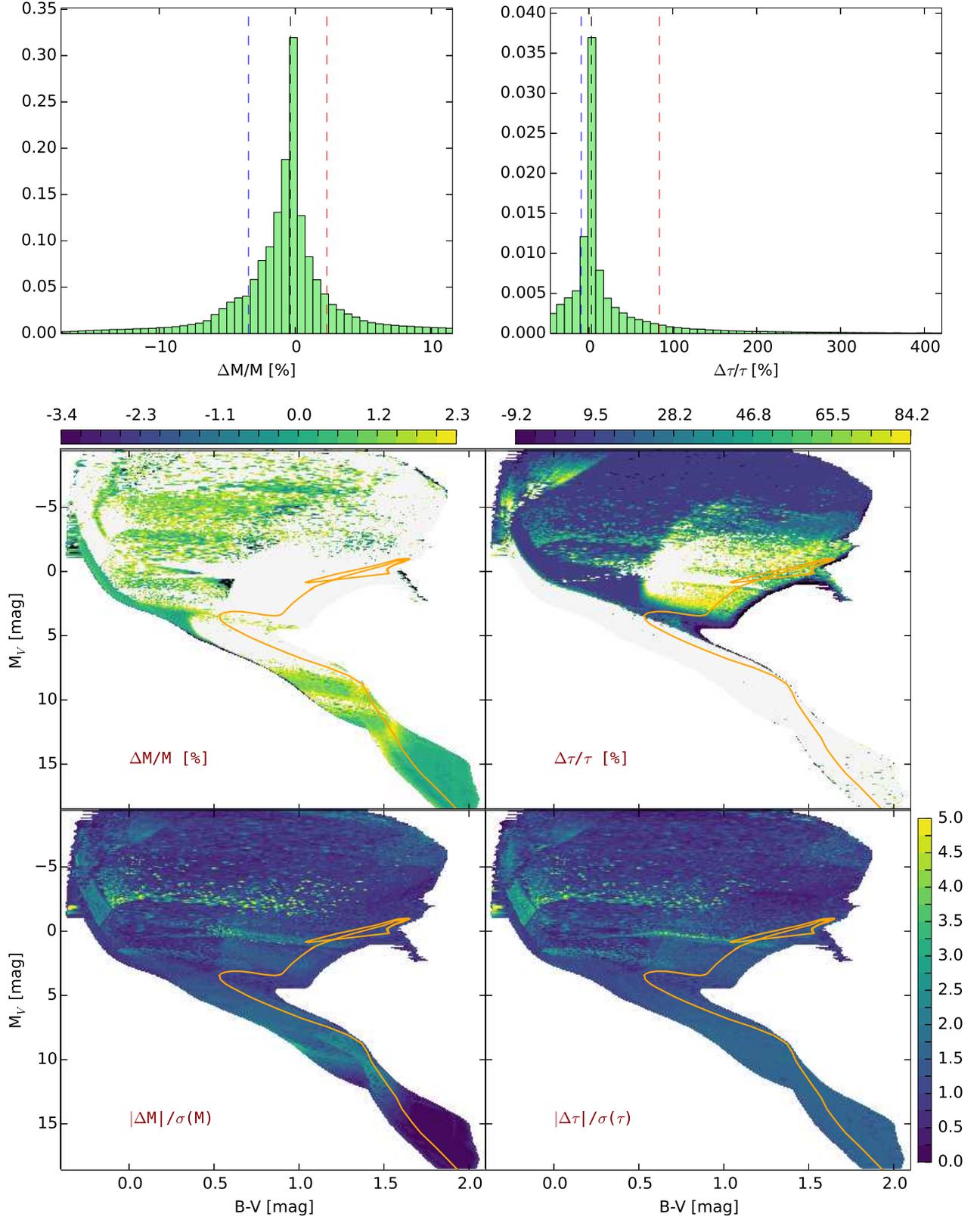}
 \caption{Histograms of the percentages $\Delta M/M$ ({\it top-left}) and $\Delta\tau/\tau$ ({\it top-right}) from the 469,941 simulations 
               generated for the median values $\sigma$([Fe/H])=0.1 dex, $\sigma(R)/R$=0.9 per cent, and $\sigma(T_\text{eff})/T_\text{eff}$=1.8 per cent, 
               corresponding to Sample I.
               Each histogram is normalized such that the integral over the full range is one. 
               Vertical lines in black, blue, and red correspond to the median and the 34th percentiles on the left and right sides, respectively. 
               Maps of $\Delta M/M$ ({\it middle-left}) and $\Delta \tau/\tau$ ({\it middle-right}), 
               on a linear scale between the 34th percentiles, against $M_\text{V}$ and $B-V$.
               Values beyond the limits are in white smoke (34th percentile on the right) and black (34th percentile on the left). 
               The orange line represents the isochrone with solar age and metallicity.
               Maps of $\lvert\Delta M\rvert$/$\sigma(M)$ ({\it bottom-left}) and $\lvert\Delta \tau\rvert$/$\sigma(\tau)$ ({\it bottom-right})
               are represented on a linear scale between 0 and 5. Values beyond the upper limit are in white.}
\label{fig:fig15}
\end{figure*}

\begin{figure}
\centering
 \includegraphics[width=85mm,angle=0]{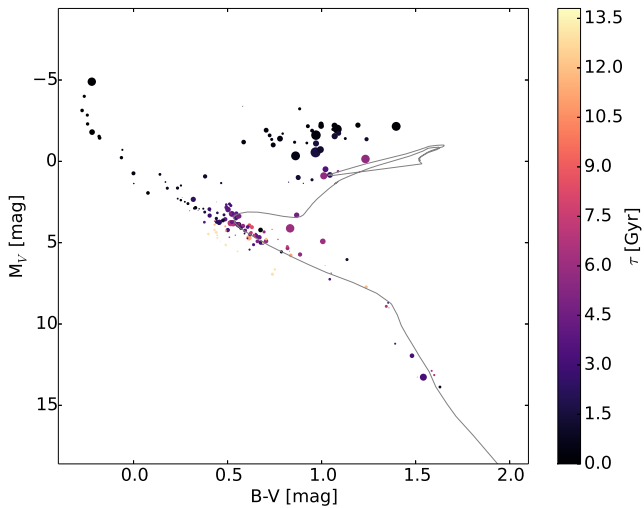}
 \caption{$M_\text{V}$ versus $B-V$ for the selected 182 DEB stars of Sample I, colour-coded by predicted age. Size of the circles linearly scales with the absolute value of the relative difference of the theoretical mass with respect to the dynamical mass. The gray line represents the isochrone with solar age and metallicity. 
}
 \label{fig:fig16}
\end{figure}

\begin{figure}
\centering
 \includegraphics[width=85mm,angle=0]{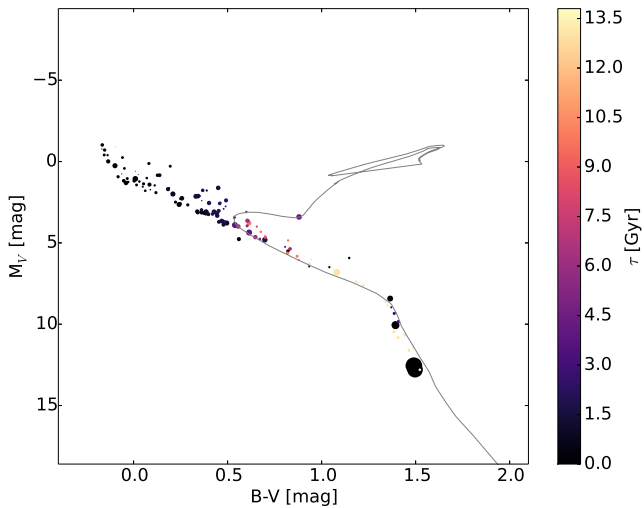}
 \caption{$M_\text{V}$ versus $B-V$ for the selected 136 DEB stars of Sample II, colour-coded by predicted age. Symbols as in Figure \ref{fig:fig17}.
}
 \label{fig:fig17}
\end{figure}

\begin{figure}
\centering
 \includegraphics[width=85mm,angle=0]{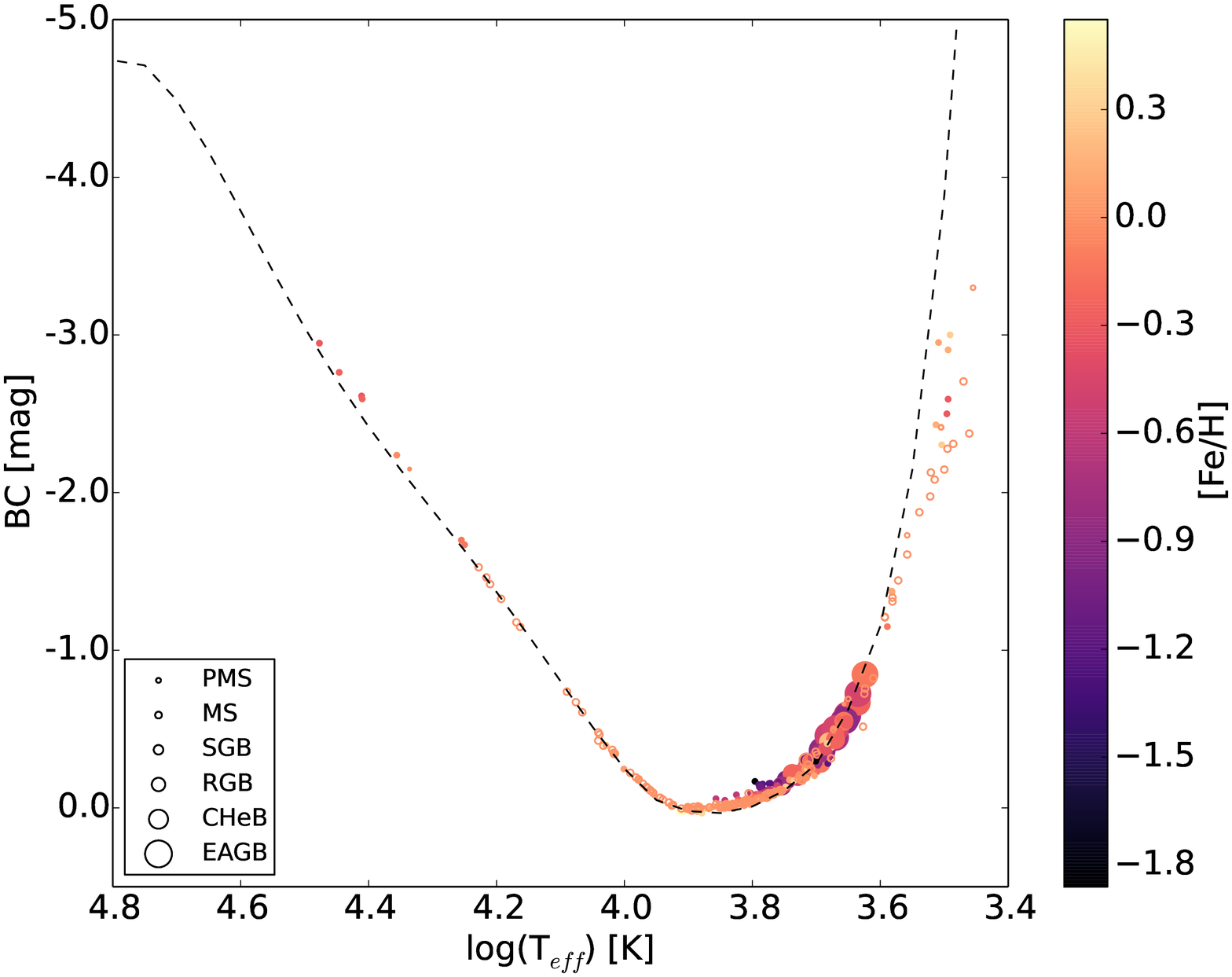}
 \caption{Bolometric correction inferred from the Bayesian method fed by {\sevensize PARSEC} v1.2S, colour-coded by metallicity. 
 Filled circles and open circles correspond to Sample I and Sample II, respectively. 
 Size of the symbols indicates the evolutionary stage, as in Fig. \ref{fig:fig12}. 
 Dashed line stands for the polynomial fits provided by \citet{torres2010b}.
}
 \label{fig:fig18}
\end{figure}

Fig. \ref{fig:fig15} displays the resulting histograms and maps of $\Delta M/M$ and $\Delta\tau/\tau$, 
and maps of $\lvert\Delta M\rvert$/$\sigma(M)$ and $\lvert\Delta\tau\rvert/\sigma(\tau)$.
Maps are built from the unstructured triangular grid of $N$ points, based on Delaunay triangulation \citep[][]{delaunay1934}.
Histograms are asymmetric, particularly that of $\Delta\tau/\tau$. 
The mean values for $\Delta M/M$ and $\Delta\tau/\tau$ are -0.36 per cent and 2.9 per cent, respectively.
The 34th percentiles on the left and right sides for the distributions of $\Delta M/M$ and $\Delta\tau/\tau$ 
are at -3.4 and 2.3, and -9.2 and 84.2 per cent, respectively. 

Maps show notable features. For instance, the horizontal branch and RGB in the $\Delta R/R$ map,
with values over the aforementioned 34th percentiles on the right side.
Also, there is a remarkable region along the main-sequence in the $\Delta\tau/\tau$ map that indicates a poor direct mean stellar age estimation due to confusion.
The $\lvert\Delta M\rvert$/$\sigma(M)$ map resembles the $\lvert\Delta\tau\rvert/\sigma(\tau)$ map.
These ratios are predominantly between 0 (i.e., no difference between the input and output values) and 1 (i.e., uncertainties equal to the differences), 
but the latter maps feature some extended regions on the loci of giants and bright giants where the absolute values of the difference between 
the inferred and input values and the uncertainties are significantly unlike. 
That said, such regions only cover small areas on the maps.

Simulations indicate that mean masses for subgiants and giants could be significantly overestimated, but uncertainties are representative of the mass deviations. 
Note we improve the situation picking up the most likely evolutionary stage (see Section \ref{getmass}). 

Our simulations indicate that mean ages can be severely biased, although uncertainties are representative of the corresponding deviations too. 
\citet{jorgensen2005} suggested to use the mode to derive ages because mean and median values tend to be biased towards intermediate ages. 
We indeed adopted the mode of the combined $G(\tau)$ function (product of the $G(\tau)$ functions of the two components corresponding to 
the most likely evolutionary stages) to estimate the stellar age of each binary system, which also reduce confusion effects. 

Figure \ref{fig:fig16} plots the CMD for the selected 182 DEB stars of Sample I, colour-coded by age and size-coded by the discrepancies in the masses.
Figure \ref{fig:fig17} displays the CMD for the 136 DEB stars of Sample II.

\subsection{Distance comparison}
\label{getdistance}

\subsubsection{Systems in Sample I}

We constrained the distances for all the 91 binary systems of Sample I (see Table \ref{tab:sampleIb}).
We used a similar approach to \citet{torres2010}, from the apparent maximum magnitude of the system, as well as the bolometric luminosity 
and the bolometric correction of each component. 
The absolute visual magnitude of the Sun employed for the calculations was $M_\text{V}$=4.81$\pm$0.03 \citep[][]{torres2010b}.
But we applied the bolometric corrections derived from our analysis. 
The resulting distances are similar (mean relative difference of $\sim$ 0.6 per cent) to those calculated from the bolometric corrections 
provided by \citet{torres2010b}, but for M dwarfs the latter approach does not work well while ours is consistent with precise distance measurements. 
For most objects, we used the apparent visual magnitude of the system $V_\text{max}$ given by DEBCat, 
but we adopted the values (with uncertainties) of \citet{graczyk2017} for the 32 DEB systems in common (15 in Sample I and 17 in Sample II). 
DEBCat does not provide uncertainties for $V_\text{max}$, so we estimate them using the relation $\sigma(V_\text{max}) = a + b~10^{(V_\text{max}-c)/5}$. 
Note $\sigma(V_\text{max})$ increases from $\sim$ 0.017 mag for $V_\text{max}\sim$ 2 mag to $\sim$ 0.7 mag for $V_\text{max}\sim$ 18 mag.

Figure \ref{fig:fig18} shows some significant discrepancies of our bolometric corrections with respect to the relation given by \citet{torres2010b}. 
Note the outliers corresponding to low mass dwarfs. 
In particular, for CM Dra we obtained a distance of 13.7$\pm$2.8\pc, 
which is in excellent agreement with the value of 14.850$\pm$0.011\pc of \textit{Gaia} DR2 \citep[][]{gaia2016,gaia2018,lindegren2018}. 
Conversely, utilising the bolometric corrections from \citet{torres2010b}, an underestimated distance of 6.7$\pm$0.7\pc is derived. 
Note we propagated all the uncertainties of the different parameters involved in the calculation of the distance,   
while uncertainties in \citet{torres2010} are underestimated.

The distances ($d$) were not corrected from extinction for nearby systems at $d<$ 150\pc.
For the rest, we performed a correction from extinction.  
We generally employed the colour excess $E(B-V)$ to calculate the absolute extinction in the $V$ band $A_\text{V}$, 
from the relationship $A_\text{V}$=$R_\text{V} E(B-V)$, assuming that $R_\text{V}$=3.10$\pm$0.05.
We utilised the Galactic dust extinction estimations of \citet{schlafly2011} for the following DEB systems: 
M4 V69, M4 V66, M4 V65, NGC 6362 V41, NGC 6362 V40, 47 Tuc V69, Kepler-35, KIC 6131659, Kepler-453, Tyc 5227-1023-1, 
V565 Lyr, KIC 7177553, Kepler-34, V568 Lyr, KIC 7037405, KIC 9970396, Kepler 1647, WOCS 40007, V375 Cep, KIC 8430105, 
NP Per, ASAS J052821+0338.5, KIC 9540226, BK Peg, BW Aqr, GX Gem, KIC 11285625, KIC 8410637, KIC 9777062, HW CMa, and V380 Cyg.
\citet{schlafly2011} determined the reddening from the differences between measured and predicted colours of stars, 
using photometric and spectroscopic observations from the Sloan Digital Sky Survey (SDSS).
For some systems, namely RW Lac, AD Boo, EK Cep\footnote{assuming an uncertainty in E(B-V) of 0.02}, 
V906 Sco, CV Vel, and V453 Cyg we chose the values of $E(B-V)$ tabulated in \citet{torres2010}.
For EW Ori, EF Aqr, and TZ For we used the reddenings given by \citet{graczyk2017}. 
\citet{stassun2016} provide $A_\text{V}$ values for CoRoT 102918586, KIC 9246715, and CoRoT 105906206, 
for which we adopted the highest value of the two uncertainties of their list.
We also employed literature estimations of $E(B-V)$ or $A_\text{V}$ for 
PTFEB 132.707+19.810 \citep[][]{taylor2006}, M55 V54 \citep[][]{kaluzny2014}, 
V785 Cep \citep[][]{meibom2009}, 
WOCS 12009 \citep[][]{taylor2007}, LV Her \citep[][]{torres2009}, CO And \citep[][]{lacy2010}, HD 187669 \citep[][]{helminiak2015}, V501 Mon \citep[][]{torres2015},  
SW Cha \citep[][]{torres2012}, and ASAS J180057-2333.8 \citep[][]{suchomska2015}.  
When uncertainties in E(B-V) are not given, we employed the median value in the sample.
Finally, we also looked for values in the literature for the extragalactic binary systems. 
In the Small Magellanic Cloud:
OGLE SMC113.3 4007 \citep[][]{graczyk2012}, 
OGLE SMC101.8 14077, OGLE SMC108.1 14904, OGLE SMC126.1 210, and OGLE SMC130.5 4296 \citep[][]{graczyk2014}.
In the Large Magellanic Cloud: OGLE-LMC-ECL-25658 \citep[][]{elgueta2016};
OGLE-LMC-ECL-03160, OGLE-LMC-ECL-06575, OGLE-LMC-ECL-09114, OGLE-LMC-ECL-09660, 
OGLE-LMC-ECL-10567, OGLE-LMC-ECL-15260, OGLE-LMC-ECL-01866, and OGLE-LMC-ECL-26122 \citep[][]{pietrzynski2013}.

\begin{table*}
\caption{Number of points N, mean $\pm$ standard deviation of relative differences $\frac{d-d_\text{L}}{d_\text{L}}$ weighted by $w_\text{1}$ ($\sigma^{-2}$; 
$\sigma$ is the uncertainty of relative difference), Pearson correlation coefficient weighted by $w_\text{1}$, slope and offset computed 
using a weighted orthogonal distance regression procedure \citep[][]{boggs1992}, and distance range for Sample I and Sample I$+$II.}
  \label{tab:groupeddeviations}
  \begin{tabular}{cccccccccc}
    \hline
    Group &   N & $\langle\frac{d-d_\text{L}}{d_\text{L}}\rangle_{\text{w}_\text{1}}$  &  $r_{\text{w}_\text{1}}$ & slope  & offset       & distance range    \\
               &      &                                                                   &                    &            & (\pcw)   & (\pcw)     \\
    \hline
Sample I                   & 70 & -0.07$\pm$0.10 &   0.987  & 0.932$\pm$0.012 & 0.09$\pm$0.06 & 1.3 -- 8095 \\
Sample I$+$II           & 133 & -0.06$\pm$0.10 & 0.987  & 0.936$\pm$0.009 & 0.09$\pm$0.05 & 1.3 -- 8095 \\
  \hline
  \end{tabular}
\end{table*}

\begin{figure}
\centering
 \includegraphics[width=85mm,angle=0]{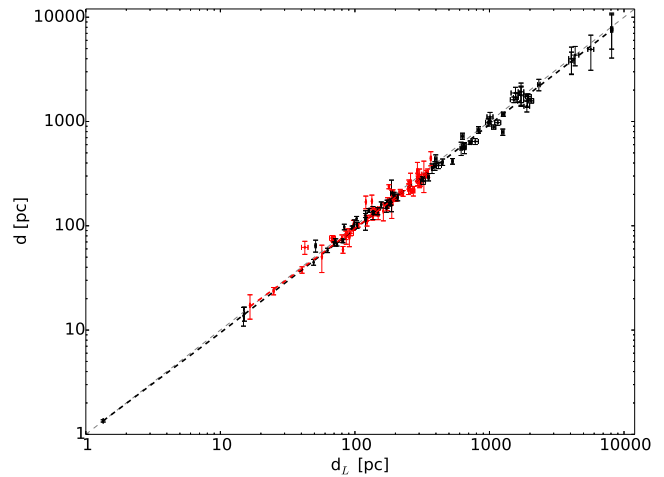}
 \caption{Distance $d$ determined from the binary parameters versus distance $d_\text{L}$ found in the literature 
 (most of them calculated from trigonometric parallaxes, all with uncertainties below 10 per cent) for 70 binary systems (black dots) 
 in Sample I and 63 systems (red dots) in Sample II.
 An underestimated extinction shifts a point above the diagonal $d$=$d_\text{L}$ and viceversa. 
 Relation 1:1 (gray dashed line) and best fitting straight lines for Sample I (in black) and Sample II (in red) 
 using the weighted orthogonal distance regression procedure of \citet{boggs1992}.
}
 \label{fig:fig19}
\end{figure}
 
There are 70 and 61 (out of the 91) systems with distances computed from trigonometric parallaxes and distance moduli 
with uncertainties below 10 and 5 per cent, respectively. 
This comparison was mostly based on parallaxes determined from \textit{Gaia} DR2 \citep[][]{gaia2016,gaia2018,lindegren2018}.
We also used \textit{Hipparcos} \cite[][]{vanleeuwen2007} parallaxes for V906 Sco and V380 Cyg.
There are other systems that belong to stellar clusters, such as V1229 Tau in the Pleiades, for which \citet{melis2014} measured 
an absolute trigonometric parallax distance of 136.2 $\pm$ 1.2 pc from very long baseline radio interferometry, which is incompatible with that derived from Hipparcos, 
but in agreement with that of \textit{Gaia} DR2 (139.5 $\pm$ 1.8 \pcw). 
We used the $\alpha$ Cen parallax updated by \citet{pourbaix2016}. 

For most of the binary systems in globular clusters and galactic clusters, 
we have derived the distances corrected from extinction using determinations of the distance moduli $\mu$ in the literature: 
M55 \citep[$\mu$= 14.13 $\pm$ 0.11 mag,][]{jimenez1998}, NGC 6362 \citep[14.74 $\pm$ 0.04 mag,][]{kaluzny2015}, 
M4 \citep[$\mu$= 12.5 $\pm$ 0.1 mag,][]{bedin2009}, 
M 67 \citep[$\mu$= 9.72 $\pm$ 0.05 mag,][]{sandquist2004},
47 Tuc \citep[$\mu$= 13.27 $\pm$ 0.14 mag,][]{zoccali2001}, 
NGC 6791 \citep[13.46 $\pm$ 0.10 mag,][]{grundahl2008}, 
NGC 188 \citep[$\mu$= 11.24 $\pm$ 0.09 mag,][]{meibom2009},
NGC 7142 \citep[$\mu$= 11.8 $\pm$ 0.5 mag,][]{straizys2014}, 
NGC 6819 \citep[$\mu$= 12.38 $\pm$ 0.04 mag,][]{brewer2016}, 
NGC 6811 \citep[$\mu$= 10.37 $\pm$ 0.03 mag,][]{sandquist2016},
and NGC 2244 \citep[$d$ = 1.6$\pm$0.2 \kpcw,][]{bonatto2009}. 
The distance determined for KIC 9777062 is $d=$ 996 $\pm$ 15\pcw, which is in excellent agreement with ours ($d=$ 1013 $\pm$ 64\pcw)
and lower than that derived from \textit{Gaia} DR2 parallax ($d=$ 1185 $\pm$ 33\pcw). 
We used the distance $d=$ 1600 $\pm$ 200 \pc for V578 Mon, instead of the more unprecise determination from \textit{Gaia} DR2 ($d=$ 1878 $\pm$ 557 \pcw).

We compared our distances calculated from the binary parameters with the literature distances for the 70 (out of 74) systems with precise 
($\sigma$ = 10 per cent) literature distances. 
Figure \ref{fig:fig19} plots the inferred distances $d$ versus those obtained from the literature ($d_\text{L}$). 
It shows a general good agreement. Differences can be mostly due to inaccuracies in the reddening and $V_\text{max}$. 
Table \ref{tab:groupeddeviations} provides the details of the best linear fit obtained from the weighted orthogonal distance regression procedure by \citet{boggs1992}.

\subsubsection{Systems in Sample II}

We also present the distances for Sample II (see Table \ref{tab:sampleIb}).
We applied no correction for extinction for systems with $d<$ 150\pc. 
For UScoCTIO 5, CV Boo, IO Aqr, HY Vir, U Oph, and all the systems in the First MOTESS-GNAT survey \citep[MG1;][]{kraus2007} apart from MG1-2056316, 
we utilised the Galactic dust extinction estimations of \citet{schlafly2011}.
For UZ Dra, VZ Cep, RZ Cha, EI Cep, and MY Cyg we used the values of E(B-V) given by \citet{graczyk2017}. 
We also employed literature values of E(B-V) for  
V1174 Ori, ZZ Uma, DM Vir, EY Cep, V442 Cyg, FS Mon, GZ CMa, V451 Oph, IQ Per, $\chi^2$ Hya, and V760 Sco \citep[][]{torres2010}, 
HP Aur, CF Tau, BF Dra, TV Nor, V596 Pup, and V335 Ser \citep[][]{stassun2016}. 
For XY Cet and V2365 Oph we adopted the values of \citet{southworth2011} and \citet{ibanoglu2008}, respectively.
For objects without uncertainties in E(B-V) we adopted the median value in the sample.

A comparison of our distances with those from the literature was performed.
We employed \textit{Gaia} DR2 parallaxes and \textit{Hipparcos} parallaxes for 63 systems with distance uncertainties below 10 per cent. 
The group with Hipparcos parallaxes is formed by ASAS J082552-1622.8, V596 Pup, $\beta$ Aur, and V1031 Ori; only $\beta$ Aur with an uncertainty below 10 per cent. 
A total of 62 (out of 66) of the systems present distance uncertainties below 5 per cent.
Figure \ref{fig:fig19} also displays the distances derived in this work for Sample II together with those of Sample I, 
illustrating the very good agreement between them and the literature values.

More precise determinations of $V_\text{max}$ and $E(B-V)$ would allow us to better constrain the distances. 
For instance, we noticed that the reddening of Kepler 1647 is likely overestimated.
Thus, better measurements are required, particularly for extragalactic DEB stars that are used as distance indicators. 
These distances are sensitive to any systematic errors in effective temperature, as a function of $T_\text{eff}^2$ \citep[][]{torres2010}.

\section{Summary and Discussion}
\label{discussion}

This work yields careful determinations of stellar parameters for 318 detached eclipsing binaries using a Bayesian method 
in combination with {\sevensize PARSEC} v1.2S stellar evolution models. 
The research is aimed at testing the accuracy and precision of this library of state-of-the-art models and the potential of the methodology.
Our approach takes as input effective temperature, radius, and and iron-to-hydrogen ratio, and their uncertainties. 
Typical precisions of $T_\text{eff}$, $R$, and [Fe/H] are 1.8 per cent, 0.9 per cent, and 0.10 dex, respectively. 
We mostly selected DEB stars from the DEBCat catalogue \citep[][]{southworth2015}, 
with additional information from \citet{torres2010} and further literature inputs.

\subsection{Input parameters: $T_\text{eff}$, $R$, and [Fe/H]}
\label{inputs}

DEBCat is not completely homogeneous. 
However, differences in the solar unit values from the literature are too small to make any significant impact on our results. 
In the worst case, the most recent determinations for the solar luminosity are lower by less than 0.5 per cent. 
We updated this stellar parameter from the radius and effective temperature.

In our exercise, uncertainties would be dominated by the precision in $T_\text{eff}$ scale.
\citet[][]{torres2010} revised the effective temperatures of their sample and used the same solar parameters. 
We showed that this catalogue is in excellent agreement with DEBCat for the objects in common.

There are different $T_\text{eff}$ scales such as that established by direct measurements 
(i.e. stellar angular diameters and bolometric fluxes) and those based on calibrated colour-temperature relations \citep[see e.g.,][]{ramirez2005}. 
The former are probably more reliable and provide an absolute $T_\text{eff}$ scale. 
In particular, the Space Telescope Imaging Spectrograph (STIS) on board Hubble Space Telescope (\textit{HST}) 
supplies absolute flux spectrophotometry with an uncertainty in the monochromatic flux at 555.75 nm (555.6 nm in air) 
of only 0.5 per cent or 0.005 mag \citep[][]{bohlin2014}, which yields accurate stellar effective temperatures 
and angular diameters \citep[see e.g.,][]{delburgo2010, allendeprieto2016, delburgo2016}.

\textit{Gaia} DR2 has brought a dramatic improvement in the astrometry available for the DEB systems considered in this paper. 
The most relevant piece of information are the parallaxes, which are as accurate as $\sim$ 20$~\mu$as, or good to 2 per cent at a distance of 1\kpc. 
This encloses the majority of the DEB stars discussed in this paper. More accurate parallaxes directly impact the inferred absolute magnitudes and the luminosity determination.

Accurate parallaxes of \textit{Gaia} will also make it possible to carry out an accurate transformation from angular diameters to stellar radii. 
The former can be derived from interferometry, but also from spectrophotometry observations, 
such as those provided by \textit{Gaia} through the BP/RP instrument in DR2. 
Thus, \textit{Gaia} is expected to yield better radii than those from interferometry, comparable to those attainable from \textit{HST} spectrophotometry.

In a next paper in this series we will determine accurate stellar properties of the STIS Next Generation Spectral Library (NGSL), 
including Bayesian predictions using {\sevensize PARSEC} v1.2S.
This work will present accurate $T_\text{eff}$ and $R$ from absolute flux spectrophotometry for stars of different spectral types.

Regarding [Fe/H], Section \ref{getmass} proves that it is suitable to assume solar metallicity to infer precise masses for nearby stars, 
although it is always better to constrain them from high-resolution spectra.

\subsection{Priors}

Our applied Bayesian method relies on a prior for the initial mass function and flat priors for age and iron-to-hydrogen ratio.
The results depend on choice of prior distributions, but this is only significant when the observations are not good enough. 
For a detailed discussion the reader is referred to \citet{jorgensen2005}.

The \textit{Gaia} data will enable to directly recover the initial mass function (IMF) from the statistics of stars in distinct kinematic 
and cluster populations in the Galaxy. 
This will be much more accurate, and less local, than the insights based on \textit{Hipparcos}, and therefore will span a wider metallicity range. 
Nevertheless, our results are mildly sensitive to the IMF.

\subsection{Mass}

From the comparison of the inferred masses with the dynamical ones (with typical precisions of 0.7 per cent), we conclude that the
{\sevensize PARSEC} v1.2S models systematically underestimate masses. 
On average, discrepancies are not so important for MS and CHeB stars, but they are significant for SGB and RGB stars, 
and even more acute for EAGB stars. 
Thus, there is a margin of improvement in the stellar evolution model library.

We also applied our analysis to the sample of giants stars (RGB, CHeB, and EAGB stars), in order to compare our results with those from \citet{ghezzi2015}. 
We employed the same fitting algorithm to their sample and Sample I, yielding consistent results, 
but our larger sample permits to show differences in the precision achieved in the determination of masses for different stellar evolutionary stages.

\subsection{Age}

We found a good agreement between our age predictions and those found in the literature, based on the comparison for 25 binary systems. 
Only two of them, CU Cnc and YY Gem, present severe discrepancies, although their membership to Castor A/B (on which is based their age estimations) is debatable. 

In general, the application of new priors or constraints on age from suitable observations is useful to reduce confusion. 
The rejection of some evolutionary stages could be also done a posteriori in view of complementary information about a particular system. 
We have only performed a posteriori conclusion on EPIC 203710387.

\subsection{Distance}

Distance accuracy is established by different parameters, such as the apparent visual magnitude. 
\textit{Gaia} will provide light curves with unprecedented photometric precision, which will allow one to determine $G_{max}$. 
This is suitable to constrain distances as performed from $V_\text{max}$. 
The estimates of interstellar extinction towards the stars in our sample will be also revised in the light of the \textit{Gaia} observations, 
but the statistical impact will be also minor in our results since reddening is modest towards most of the eclipsing systems considered for this research. 
On the other hand, these new values are expected to be fully consistent with the effective temperatures. 

We proved that bolometric corrections must be accurate enough to properly derive distances from the approach described in Section \ref{getdistance}. 
Calibrated relations such as those provided by \citet{torres2010b} are not good enough for low-mass dwarfs, as we illustrate with the case of CM Dra. 
Bolometric corrections were derived from our Bayesian approach. 
We utilised them to arrive at distances that are consistent with accurate distances from the literature for the full range of masses.
%This comparison could be updated from new \textit{Gaia} data releases. 
It is worth nothing that accurate distances can help to further constrain the stellar parameter's solutions. 
We will discuss this in a forthcoming paper.

\subsection{Scope of this research}

This is a complete study on the accuracy and precision of the {\sevensize PARSEC} v1.2S stellar evolution models carried out from the comparison 
with reliable parameters of DEB stars. 
All the tests and evaluations described in this paper were performed making use of this particular library of stellar evolution models.
However, we expect the fundamental limitations associated with the statistical nature of the method to recover stellar masses will be fairly similar 
for other stellar evolution models. 
Anyway, a fair comparison of those stellar evolution models publicly available would be quite interesting. 
Despite the physics adopted by different modellers are similar, the high accuracy and wide range of the dynamical masses for 
the components of DEB stars can reveal what the optimal choices for some of the modelling parameters are.

It is also worth noting that significant efforts have been made to compile accurate stellar parameters of DEB stars, 
but these compilations do not sample the full parameter space, so we can only partly test any stellar evolution model library.
In this paper, we release inferred ages and distances for the most extensive compilation of DEB stars with accurate parameters performed so far. 
We will eventually update our results with new measurements for well-known DEB systems at different stellar evolutionary stages.

In summary, there is still much work to do to expand the number of DEB systems with accurate information and to apply the same tests in this paper 
to other state-of-the-art stellar evolution codes.

\section{Conclusions}
\label{conclusions}
The present work is an endeavour to properly infer and release parameters from stellar evolution models for a selected sample of 318 well-known 
detached eclipsing binaries.

We conclude that the adopted Bayesian method in combination with {\sevensize PARSEC} v1.2S stellar evolution models employed for this research, 
taking on input accurate $R$, $T_\text{eff}$, and [Fe/H], and their uncertainties for the sample, 
yield masses that are affected by a systematic offset and a dispersion that depends on the evolutionary stage, 
with significant discrepancies for sub-giant and giant stars, and more sharply for EAGB stars.
It is plausible to assume solar metallicities for stars closer than 300\pc as we prove using a control sample of DEB stars with known metallicities.
We also arrive at distances and ages that are in good agreement with reliable values in the literature, at least for the examined cases. 
Bolometric corrections must be carefully determined in order to obtain precise distances, as evident for the early spectral-type pair CM Dra.

\section*{Acknowledgements}

This work has been supported by Mexican CONACyT research grant CB-2012-183007.
CAP is thankful to the Spanish MINECO for support through grant AYA2014-56359-P. 
This research has made use of the SIMBAD database, operated at CDS, Strasbourg, France and NASA's Astrophysics Data System.
This research has made use of the VizieR catalogue access tool, CDS, Strasbourg, France. The original description of the VizieR service was published in A\&AS 143, 23.
This research has made use of the NASA/ IPAC Infrared Science Archive, which is operated by the Jet Propulsion Laboratory, California Institute of Technology, under contract with the National Aeronautics and Space Administration.
This work has made use of data from the European Space Agency (ESA) mission {\it Gaia} (\url{https://www.cosmos.esa.int/gaia}), 
processed by the {\it Gaia} Data Processing and Analysis Consortium (DPAC, \url{https://www.cosmos.esa.int/web/gaia/dpac/consortium}). 
Funding for the DPAC has been provided by national institutions, in particular the institutions participating in the {\it Gaia} Multilateral Agreement.

%%%%%%%%%%%%%%%%%%%%%%%%%%%%%%%%%%%%%%%%%%%%%%%%%%

%%%%%%%%%%%%%%%%% APPENDICES %%%%%%%%%%%%%%%%%%%%%
\appendix

\section{Interpolated grid}
\label{appendix:grid}

\begin{figure*}
\centering
 \includegraphics[width=175mm,angle=0]{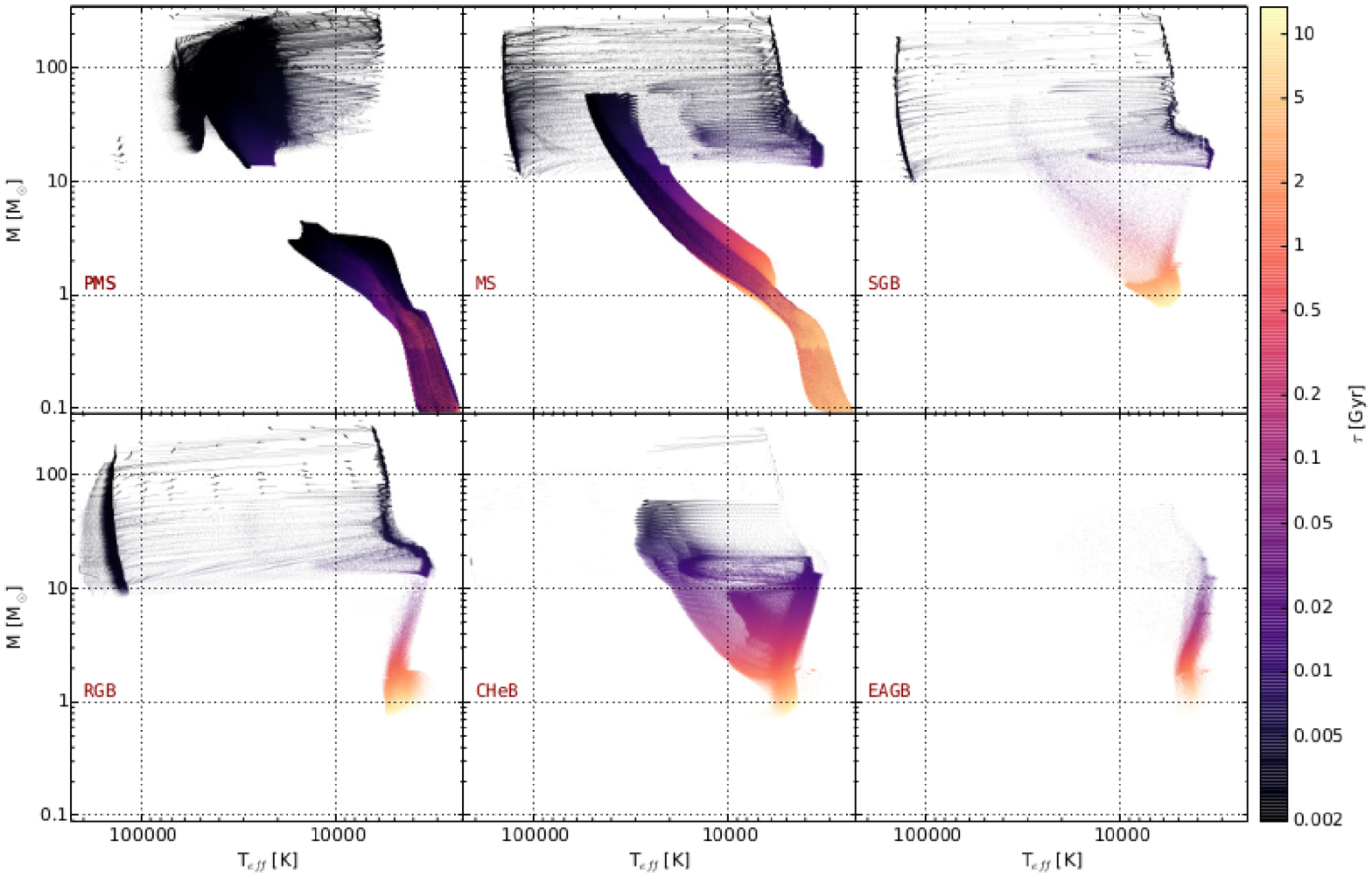}
 \caption{Loci of the {\sevensize PARSEC} v1.2S models, colour-coded by age, 
               in the $T_\text{eff}$--$M$ diagram for the stellar evolutionary stages PMS, MS, SGB, RGB, CHeB, and EAGB.
               Each point represents a model from the original values in the library after regridding. 
}
 \label{fig:appx1}
\end{figure*}

Figure \ref{fig:appx1} displays age against $T_\text{eff}$ and $M$ from the library of {\sevensize PARSEC} v1.2S stellar evolution models after regridding. 
As already mentioned in Section \ref{evolutionmodels}, there are 61,227,575 models (39,493,145 when excluding PMS phase) in the interpolated grid, 
many more than the number of original models.
The distribution of evolutionary stages significantly changes with respect to the original grid shown in Figure \ref{fig:fig1}.   
Importantly, DEB stars of Sample I and those of Sample II are in positions of the diagram that are generally better sampled by the interpolated grid.

\section{Stellar evolution}
\label{appendix:evolution}

\begin{figure}
\centering
 \includegraphics[width=85mm,angle=0]{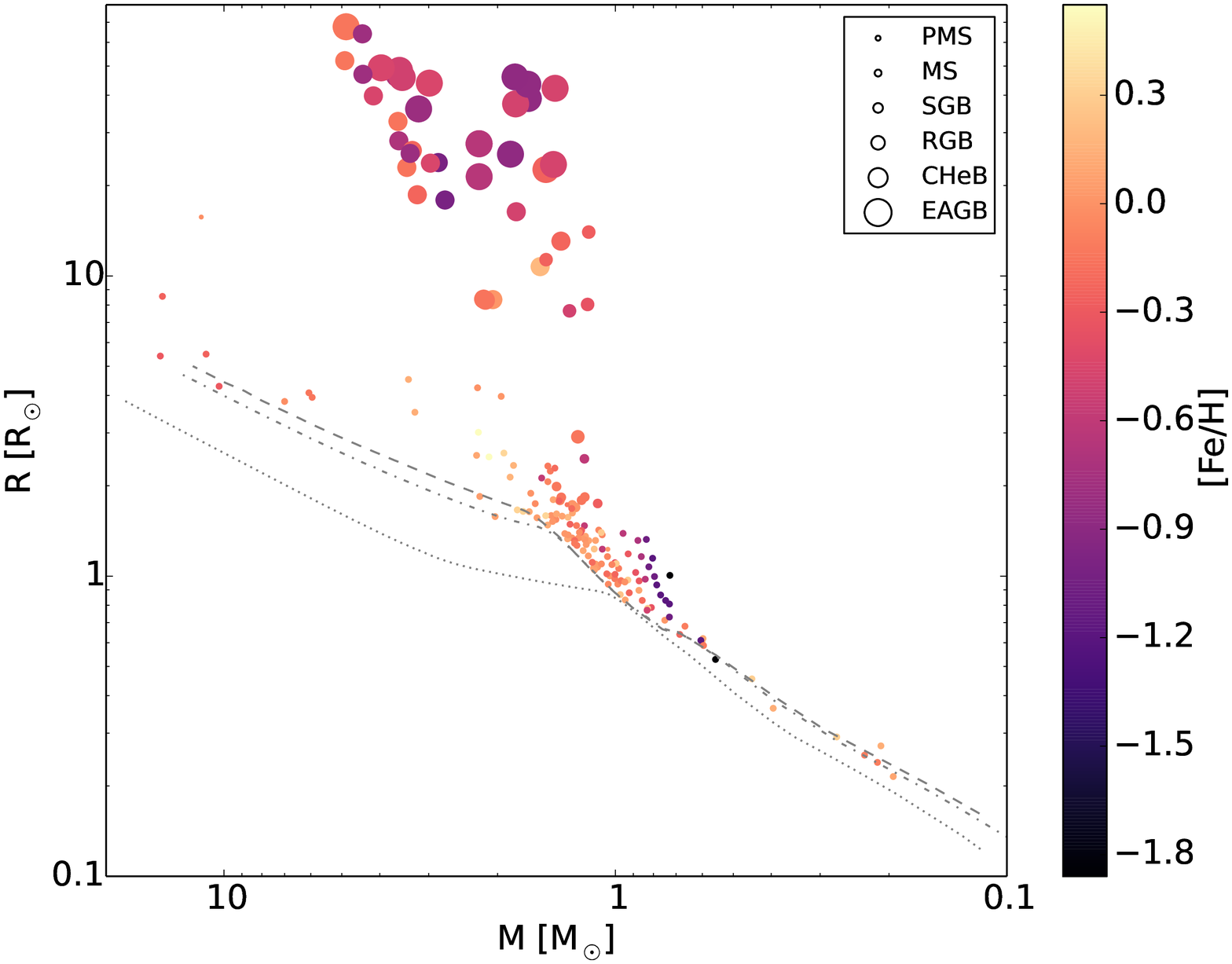}
 \caption{Measured radius versus dynamical mass for stars in Sample I, colour-coded by metallicity. 
 Lines as in Fig. \ref{fig:fig9}.
}
 \label{fig:appx2}
\end{figure}

\begin{figure}
\centering
 \includegraphics[width=85mm,angle=0]{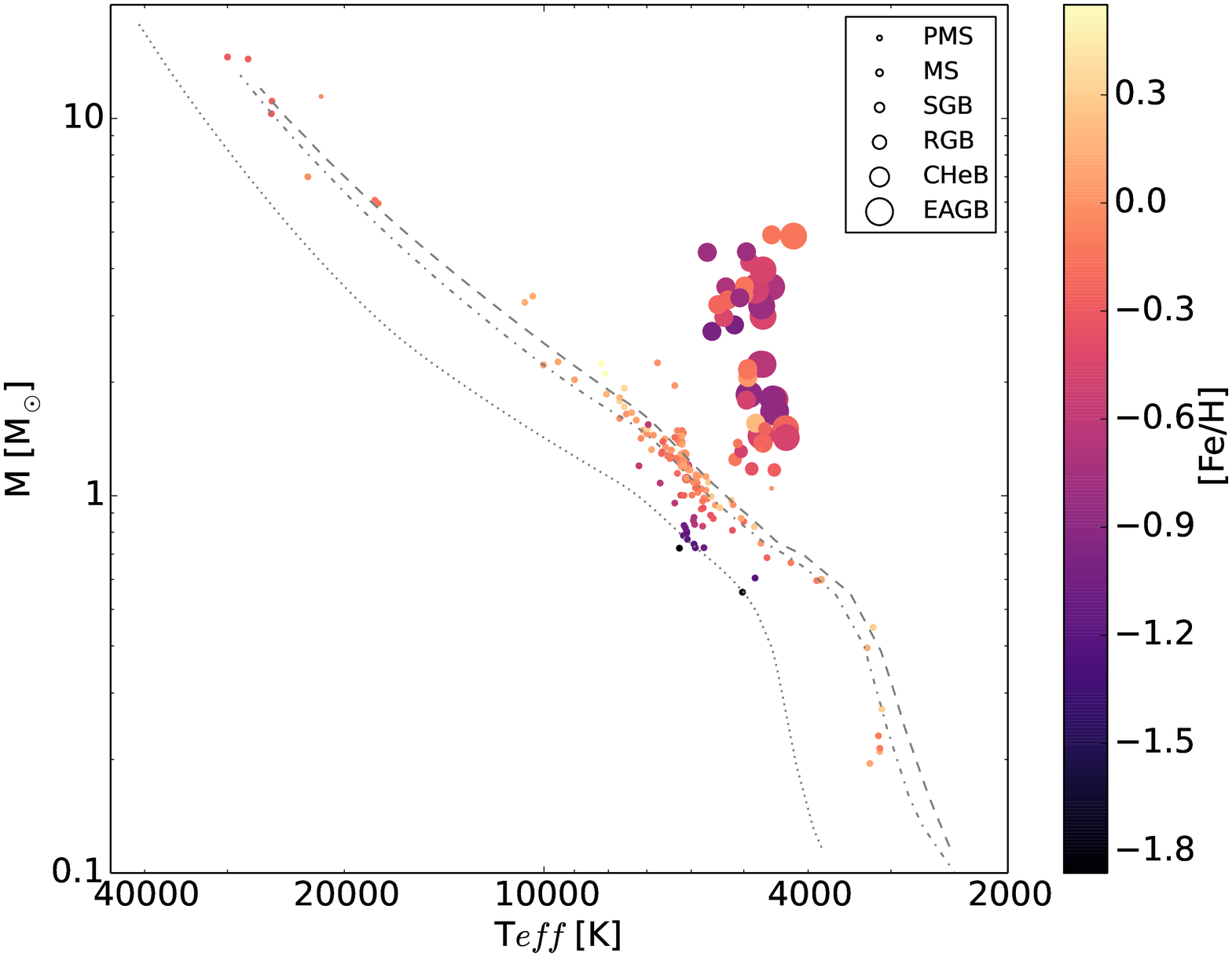}
 \caption{Dynamical mass versus effective temperature for stars in Sample I, colour-coded by metallicity. 
 Lines as in Fig. \ref{fig:fig9}.
}
 \label{fig:appx3}
\end{figure}

\begin{figure}
\centering
 \includegraphics[width=85mm,angle=0]{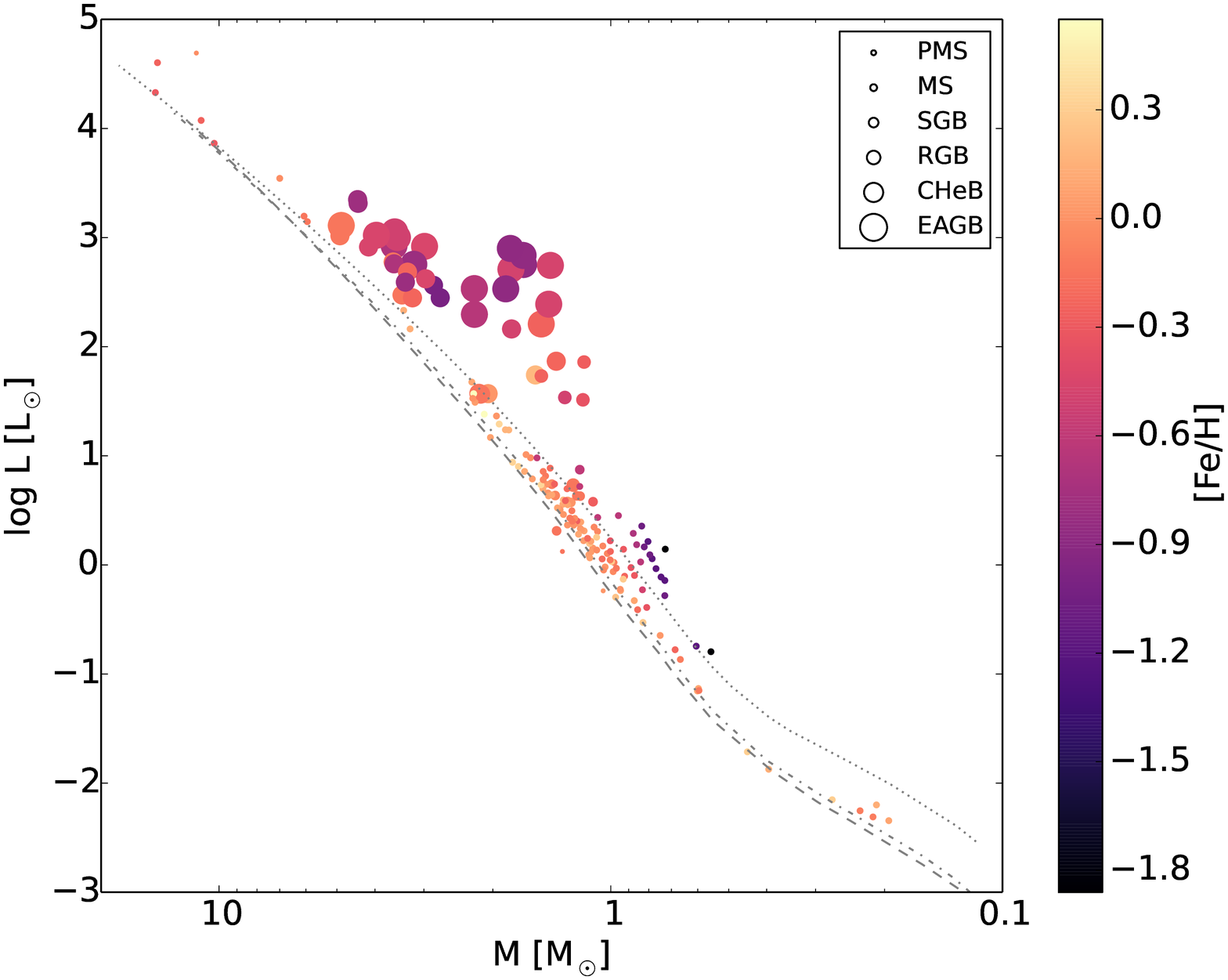}
 \caption{Measured luminosity versus dynamical mass for stars in Sample I, colour-coded by metallicity. 
 Lines as in Fig. \ref{fig:fig9}.
}
 \label{fig:appx4}
\end{figure}

Figures \ref{fig:appx2}, \ref{fig:appx3}, and \ref{fig:appx4} respectively show the diagrams $R-M$, $M-T_\text{eff}$, and $L-M$ for Sample I, 
which are useful to interpret the evolutionary effects outlined in Section \ref{stellarevolutioneffects} and also illustrated in Figure \ref{fig:fig9}.
We graphically include information about the metallicity and stellar evolutionary stage of the binaries,
as well as the zero-age-main-sequence loci for the solar and the two most extreme metallicities in the {\sevensize PARSEC} v1.2S models 
(i.e., [Fe/H]=-2.1, 0.0 and 0.4). Main-sequence stars of different metallicity are grouped in the four diagrams. 
Low-metallicity MS binary stars are located to the right of higher metallicity MS stars in the $R-M$ and $L-M$ diagrams,
and below in the $R-T_\text{eff}$ and $M-T_\text{eff}$ diagrams.

\begin{figure}
\centering
 \includegraphics[width=85mm,angle=0]{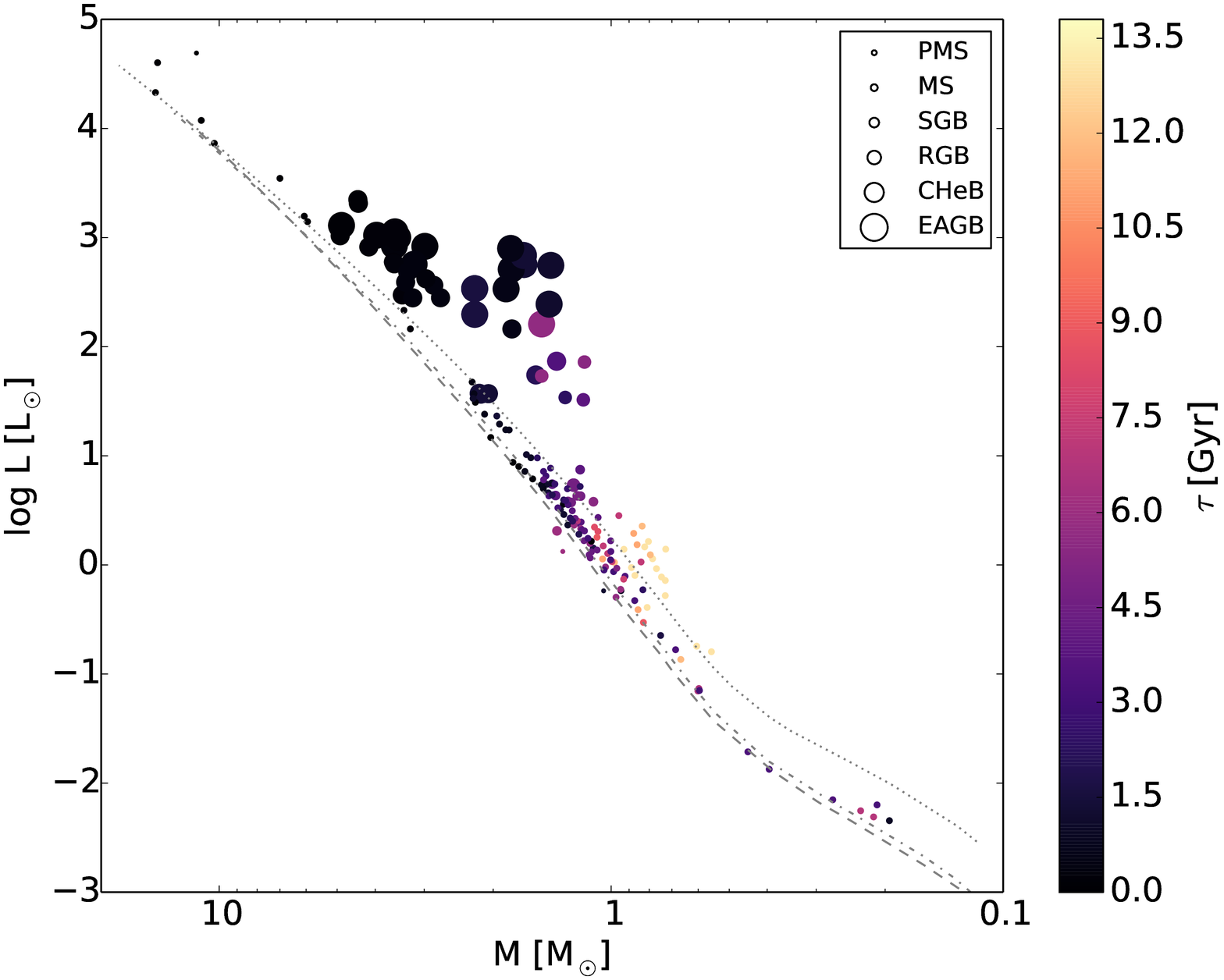}
 \caption{Measured luminosity versus dynamical mass for stars in Sample I, colour-coded by predicted age. 
 Lines as in Fig. \ref{fig:fig9}.
}
 \label{fig:appx5}
\end{figure}

\begin{figure}
\centering
 \includegraphics[width=85mm,angle=0]{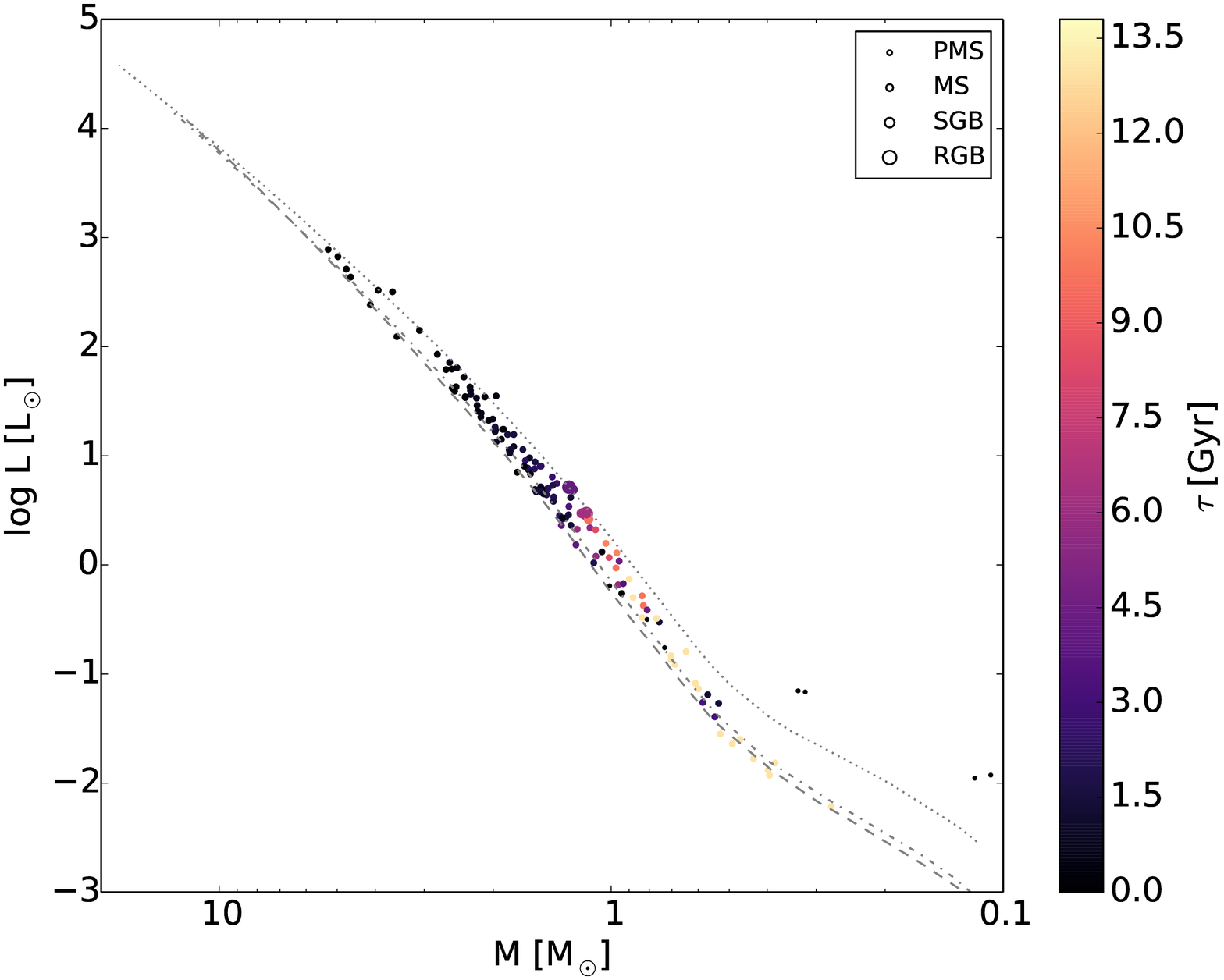}
 \caption{Measured luminosity versus dynamical mass for stars in Sample II, colour-coded by predicted age. 
 Lines as in Fig. \ref{fig:fig9}.
}
 \label{fig:appx6}
\end{figure}

Figure \ref{fig:appx5} displays the same diagram as Figure \ref{fig:appx4}, but colour-coded by age, as inferred in this work.
The MS stars with masses higher than 0.8 M$_{\sun}$ that are enclosed by the ZAMS for [Fe/H] = -2.1 and 0.4 are younger towards higher luminosities.
We also find that the low metallicity MS binaries we mentioned before are the oldest in the sample while the giants are generally young.
Figure \ref{fig:appx6} shows the same diagram as Figure \ref{fig:appx5} but for the nearby DEB stars with assumed solar metallicities (Sample II).

\section{Likelihood function}
\label{appendix:likelihood}

\begin{figure}
\centering
 \includegraphics[width=85mm,angle=0]{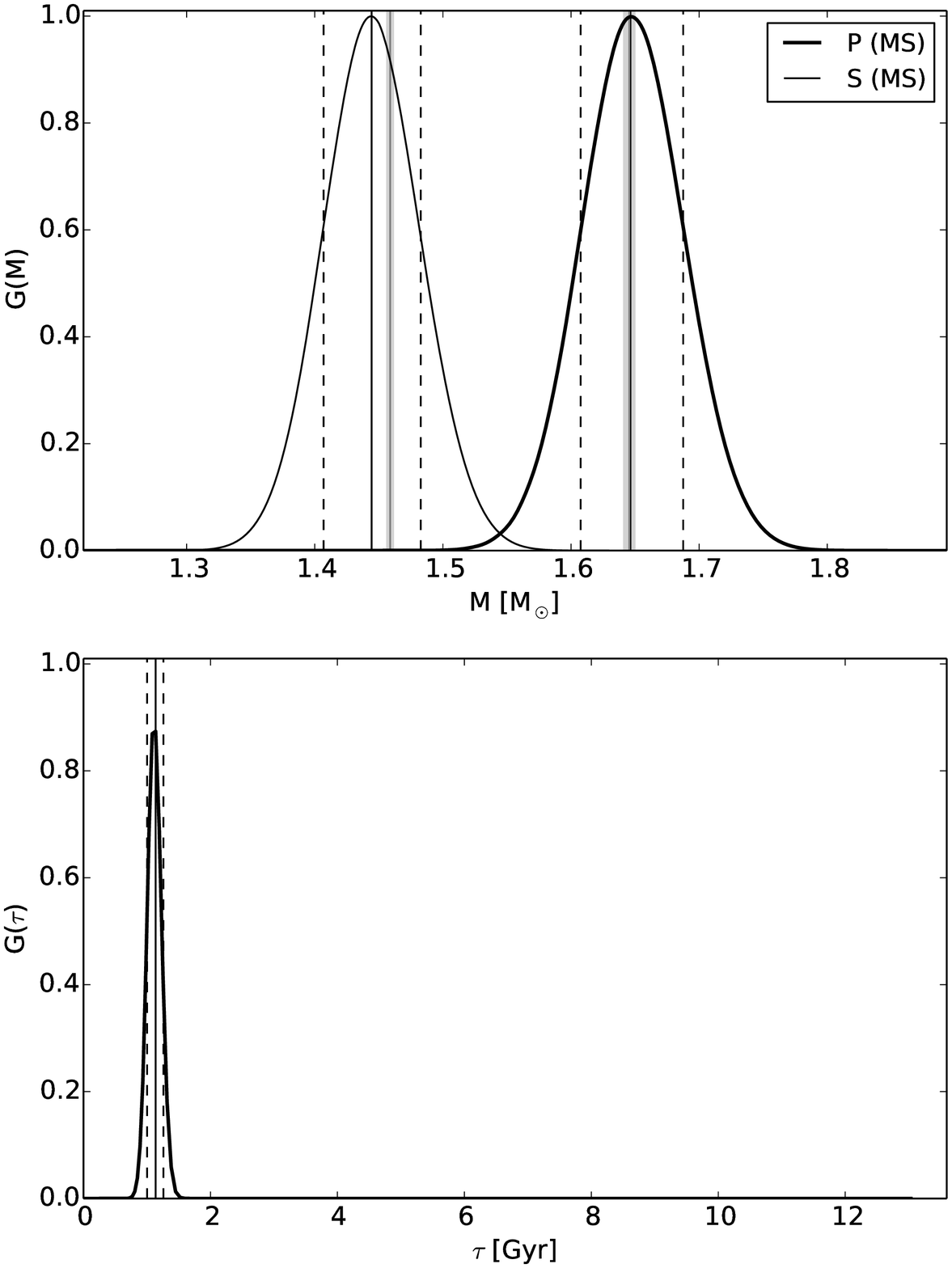}
 \caption{Posterior probability density functions $G(M)$ (top) of the main-sequence primary (thick line) and secondary DEB binaries of V501 Mon. 
 Modes (black continuous lines) and corresponding 68th percentiles (dashed lines) together with the dynamical masses, 
 which are represented by gray boxes (widths are equal to uncertainties) are shown. 
 Combined $G(\tau)$ (bottom) of this system, with the mode (continuous line) and 68th percentiles (dashed lines). 
}
 \label{fig:appx7}
\end{figure}

Figure \ref{fig:appx7} displays the posterior probability density functions $G(M)$ of the DEB binaries in the system V501 Mon (upper panel). Also, the combined posterior probability density function $G(\tau)$ (lower panel).

\section{Theoretical ZAMS}
\label{appendix:zams}

\begin{table*}
  \caption{Fitting coefficients $a_k$ of relationships $\log R$ vs. $\log M$ and $\log L$ vs. $\log M$ for ZAMS with [Fe/H]=-2.1, 0.0, and 0.4.}
  \label{tab:fitting}
  \begin{tabular}{crrrrr}
   \hline
                & \multicolumn{5}{c}{ZAMS with [Fe/H]=-2.1}\\
                & \multicolumn{3}{c}{$\log R$ vs. $\log M$} & \multicolumn{2}{c}{$\log L$ vs. $\log M$} \\
    $a_k$ & 0.116$<M<$0.746 M$_{\sun}$ & 0.746$\leq M \leq$2.37 M$_{\sun}$ & 2.37$<M<$26.5 M$_{\sun}$ & 0.116$<M<$0.834 M$_{\sun}$ & 0.834$\leq M<$26.5 M$_{\sun}$ \\
   \hline
     $a_0$      & -0.16216446661 & -0.0748888821792  & 0.0904007216371 & 0.190995504924 & 0.244049226369  \\
     $a_1$	& -0.38805089363 & 0.67915099344     & -0.89617266172  & 3.26878126554  & 4.47814358163   \\
     $a_2$	& -7.64472974862 & -3.7721850171     & 2.56640563428   & -9.50819474316 & -1.29894177262  \\
     $a_3$	& -17.4426838134 & 3.98869641645     & -1.83951204181  & -13.4720011902 & 0.207228675633  \\
     $a_4$	& -16.7949984035 & 63.4800494793     & 0.489278496927  & 68.1643134593  & 0.41816263001   \\
     $a_5$	& -5.80766674243 & -240.157554021    &                 & 198.691937731  & -0.222040272357 \\
     $a_6$	&                & 249.47681417      &                 & 190.871355862  &                 \\
     $a_7$      &                &                   &                 & 64.0636779804  &                 \\
   \hline
                & \multicolumn{5}{c}{ZAMS with [Fe/H]=0} \\
                & \multicolumn{3}{c}{$\log R$ vs. $\log M$} & \multicolumn{2}{c}{$\log L$ vs. $\log M$} \\
    $a_k$ & 0.098$<M<$0.738 M$_{\sun}$ & 0.738$\leq M \leq$1.515 M$_{\sun}$ & 1.515$<M<$13.7 M$_{\sun}$ & 0.098$<M<$0.818 M$_{\sun}$ & 0.818$\leq M<$13.7 M$_{\sun}$ \\
   \hline
     $a_0$      & -0.247156304987 & -0.0555786069013 & 0.126500320173  & -0.306003120799 & -0.149688466976 \\
     $a_1$	& -1.70127232474  & 1.18511313356    & 0.0016887876638 & 0.719361241361  & 4.8483356011    \\
     $a_2$	& -11.3928643593  & 1.60851647111    & 1.09696529771   & -34.8050273637  & 0.871576730532  \\
     $a_3$	& -21.8477584047  & -4.74542772695   & -1.11864011559  & -138.023372264  & -13.9137197785  \\
     $a_4$	& -18.7380937735  & -10.4954160139   & 0.649407795892  & -261.16922125   & 36.4651454898   \\
     $a_5$	& -5.95966410638  & -31.147277278    & -0.155004112888 & -270.705451791  & -45.3618694221  \\
     $a_6$	&       	  & -260.92862205    &  	       & -148.864184389  & 27.6999758074   \\
     $a_7$      &                &                   &                 & -33.8788988121  & -6.68690051555  \\
   \hline
                & \multicolumn{5}{c}{ZAMS with [Fe/H]=0.4}  \\
                & \multicolumn{3}{c}{$\log R$ vs. $\log M$} & \multicolumn{2}{c}{$\log L$ vs. $\log M$} \\
    $a_k$ & 0.117$<M<$0.749 M$_{\sun}$ & 0.749$\leq M \leq$1.760 M$_{\sun}$ & 1.760$<M<$14.2 M$_{\sun}$ & 0.117$<M<$0.815 M$_{\sun}$ & 0.815$\leq M<$14.2 M$_{\sun}$ \\
   \hline
     $a_0$      & -0.257199694041 & -0.0559461944729 & 0.160198268047  & -0.630072871576 & -0.253013181746 \\
     $a_1$	& -1.74741688396  & 1.14944048604    & -0.0133762020281& -5.6330458262	 & 4.88721632146   \\
     $a_2$	& -11.319495835   & 1.24120986216    & 1.64941444852   & -101.665424716  & 1.37790417536   \\
     $a_3$	& -21.7394423921  & 3.21426232384    & -2.62385062185  & -480.153852051  & -17.1628010294  \\
     $a_4$	& -18.855859093   & -7.51653429715   & 2.14351000165   & -1187.57475105  & 46.7534510249   \\
     $a_5$	& -6.10972714725  & -273.533834478   & -0.667634163833 & -1629.47516106  & -60.7181241173  \\
     $a_6$	& 	          & 672.924046386    & 		       & -1168.24185724  & 38.5558578279   \\
     $a_7$      &                 &                  &                 & -340.099861567  & -9.64117512452  \\
   \hline
  \end{tabular}
 \end{table*}

We performed polynomial fittings to the theoretical zero-age main-sequence (ZAMS) $\log R$ vs. $\log M$ relationships for [Fe/H]=-2.1, 0.0, and 0.4 
from {\sevensize PARSEC} v1.2S stellar evolution models. The polynomial fits are expressed as: 
\begin{equation}
\log R = \sum_{k=0}^{n} a_k (\log M)^k
\end{equation}

We distinguished between different mass ranges.

Similarly, we arrived at the coefficients $a_k$ that fit the ZAMS $\log L$ vs. $\log M$ relationships for [Fe/H]=-2.1, 0.0, and 0.4:
\begin{equation}
\log L = \sum_{k=0}^{n} a_k (\log M)^k
\end{equation}

Table \ref{tab:fitting} lists the results.

\bsp	% typesetting comment
\label{lastpage}

\newpage
%The following tables should be only given in the electronic version. 
%Here we show them in order the referee can see the results.

\end{document}